\documentclass[11pt]{article}

\usepackage[pdftex]{graphicx}

\usepackage{amsmath} 
\usepackage{amssymb}
\usepackage{overpic}
\usepackage{fancyvrb}

\begin{document}
\VerbatimFootnotes

\thispagestyle{empty}

\noindent
\textbf{\Large Deriving on-shell open string field amplitudes
}

\vspace{3mm}

\noindent
\textbf{\Large without using Feynman rules}

\vspace{10mm}

\noindent
{
Masuda, Toru$^{a,\,}$\footnote[1]
{e-mail: masudatoru@gmail.com}
\qquad 
Matsunaga, Hiroaki$^{a,\,b,\,}$\footnote[2]
{e-mail: matsunaga@math.cas.cz
}

\vspace{5mm}

\noindent
$^a$CEICO, Institute of Physics, the Czech Academy of Sciences\\
\textit{Na Slovance 2, 18221 Prague 8, Czech Republic}\\

\noindent
$^b$Institute of Mathematics, the  Czech Academy of Sciences\\
\textit{Zinta 25, 11567 Prague 1,  Czech Republic}\\

\vspace{20mm}
\noindent
Abstract: 
We present a series of new gauge invariant quantities in Witten's open string field theory. They are defined for a given set of open string states which satisfy the physical state condition  around a classical solution.
For known classical solutions, we show that these gauge invariant quantities compute on-shell tree-level scattering amplitudes around the correspondent D-brane configuration. \\

\clearpage
\tableofcontents

\setcounter{page}{1}
\section{Introduction}

\noindent
Feynman's formulation of quantum field theory immensely simplified the perturbative calculation of physical quantities. 
His diagramatic method appeared in 
the late 1940s, and exerted its power on the central problems of the time. 
While the interests of the community have partially shifted to non-perturbative aspects, 
Feynman's methods are still a fundamental tool in quantum field theory. 
\\

\noindent
In general, a string field theory is formulated %
so that Feynman's method works appropriately. It is thus by definition a proper way of doing the perturbative calculation;  
however, it would be natural to imagine that there exists an alternative way peculiar to the string field theory, as the degrees of freedom are now replaced by strings whose trajectory is a two dimensional surface.

If such an alternative method was developed, 
it would provide a new paradigm in perturbative calculations. It might 
(1) provide us a new interpretation of perturbative calculation alternative to Feynman diagrams, 
(2) give us a new way to decompose the moduli space of the Riemann surfaces and 
(3) enable calculations around other backgrounds or D-brane configurations apart from the one we used to define the theory. 
 \\

\noindent
We have a prototypical example in mind: Ellwood's interpretation of the gauge invariant observable in open cubic string field theory \cite{Witten}. 
The well-known relation given in \cite{Ellwood} can be viewed as a formula to calculate the on-shell closed string tadpole amplitude 
around a new D-brane configuration, which is represented by the classical solution 
(see the end of Section 2.3 for more details). \\

\noindent
Following Ellwood's work, we  would like to construct a gauge invariant quantity corresponding to the on-shell tree-level amplitude.  
Our strategy is quite simple: 
\begin{enumerate}
\item Take some natural objects of the theory.
\item Find a combination of these natural objects which is invariant under 
\begin{enumerate}
\item the gauge transformation of the classical solution. 
\item the linearized gauge transformations of the external states, which implies decoupling of the null states.
\end{enumerate}
\end{enumerate}

\noindent
In the first quantized string theory, 
decoupling of the null states is essential to obtain a correct expression for scattering amplitudes~\cite{Polchinski:1998rq}. 
Also, in covariant string field theory, 
space-time gauge invariance is an important guideline to achieve 
the  modular invariance of the amplitudes \cite{GMW, Z2}. 
It is therefore quite natural to look for a formula for on-shell amplitudes
by postulating these fundamental requirements. 

In addition,  we will use a tachyon vacuum solution as a basic building block of our formula. 
We also use an extra physical input, namely  triviality of the cohomology around the tachyon vacuum solution. 
In this sense, the present work was possible thanks to the recent development of analytic methods. \\

\noindent
In the next section, we will explain our conventions as well as some basic materials. We then introduce a definition of the new gauge invariant quantities in Sections 3 and 4 
and also show properties of these quantities including the gauge invariance and decoupling of the null states. 
In Section~5 we will show that they match the on-shell amplitude using known solutions.

The present paper is, in some aspects, a lead-up to possible subsequent works, for there are many important  topics left untouched. 
In Section 6 we will describe and further elaborate them to conclude this work.
\\

\section{Summary of basic materials}
In this section, we will  provide a brief explanation of the open string field theory 
for a wider range of readers, 
while it will not be  a self-contained review.
We cite a lecture note \cite{Taylor:2003gn} as an accessible reference covering the subjects of Sections 2.1 and 2.2.
For full description on the subjects of Section~2.3, see reviews \cite{reviews_analytic_solution_0, reviews_analytic_solution_1, reviews_analytic_solution_2} or references \cite{Schnabl,Okawa,Schnabl:2002gg,Rastelli:2000iu}.

\subsection{Algebraic structure}
In this paper, we consider an open bosonic string field theory with the Chern-Simons type action, which was first written down by Witten \cite{Witten}:
\begin{equation}
\label{eq:action}
S[\Phi]=-\frac{1}{g_o^2}\left(\frac{1}{2}\int \Phi\ast Q\Phi+\frac{1}{3}\int \Phi \ast\Phi\ast\Phi\right). 
\end{equation}
Here  
$\Phi$ is an open string field, which is an element of some graded vector space
 $\mathcal H$. We assume $\mathcal H$ is equipped with an associative product $\ast$ (a star product) and a differential $Q$, and $(\mathcal H, \ast, Q)$ is a differential graded algebra. 
This means, the grading is additive with $\ast$, and  $Q$ is  a derivation of the star product with degree one, which satisfies $Q^2=0$. 
The Witten integral $\int$ is a linear map from $\mathcal H$ to $\mathbb C$. 
In addittion, we also postulate 
\begin{equation}
\int Q \Phi=0 \quad \text{for} \quad \forall \Phi\in \mathcal H,
\end{equation}
 and the cyclic property of the 2-vertex
\begin{equation}
\int \Phi_1\ast \Phi_2=(-1)^{|\Phi_1||\Phi_2|}\int \Phi_2\ast \Phi_1, 
\end{equation}
where $|\Phi|$ denotes the degree of $\Phi$. 
With these relations,  the action $S[\Phi]$ is invariant under the gauge transformation 
\begin{equation}
\label{eq:futyftuyfuty}
\Phi\to \Phi'=U(Q+\Phi)U^{-1}, 
\end{equation}
where $U$ and $U^{-1}$ are elements of $\mathcal H$ at degree zero.
As we see, the structure of this theory resembles that of the Chern-Simons theory, except for the fact there exist elements with negative degrees in~$\mathcal H$. 
\\

\noindent
The equation of motion derived from \eqref{eq:action} is 
\begin{equation}
Q\Phi+\Phi\ast\Phi=0.
\end{equation}
In the following, we will exclusively use the letter $\Psi$ with an 
optional suffix to denote a classical solution to the equation of motion. 
We can expand the action \eqref{eq:action} around $\Psi$ as follows:
\begin{equation}
\label{eq:tenkaishitasayou}
S[\Psi+\Phi]=-\frac{1}{g_o^2}\left(\frac{1}{2}\int \Phi\ast Q_\Psi\Phi+\frac{1}{3}\int \Phi \ast\Phi\ast\Phi\right)+e(\Psi),
\end{equation}
where $Q_\Psi$ is a nilsquare operator given by
\begin{equation}
Q_\Psi \phi=Q \phi+\Psi *\phi-(-1)^{|\Psi||\phi|}\phi*\Psi,
\end{equation}
and $e(\Psi)$ is energy of $\Psi$, which is a number.\\

\noindent
Note that the difference of two classical solutions $(\Psi_2-\Psi_1)\equiv \Psi_{21}$ satisfies the equation of motion around the classical solution $\Psi_1$:
\begin{equation}
\label{eq:eomoemoe}
Q_{\Psi_1}\Psi_{21}+\Psi_{21}\ast\Psi_{21}=0.
\end{equation}

\noindent
To this point, we have only focused on the algebraic framework of the theory. 
In the next subsection, we will 
see that 
this algebraic framework can be realized 
as a field theory of open strings.

\subsection{Realization of the theory using BCFT}

\noindent
Remember that the dynamical variable of quantum field theory is a field, which is a function of the position of corresponding particle. 
Similarly, we regard {an open string field} $\Phi$ as a function of configurations of an open string. 
%
%
Witten proposed that 
we can construct a theory satisfying the above algebraic structure with open string fields (in the above sense) under a concrete interpretation of $\ast$ and $\int$ (see (b), (c), (d) in Figure 1).\footnote{
\label{foot:leftright}
We use \textit{the right handed convention} for the star product: 
 $\psi_1\ast \psi_2$ means that the right half of $\psi_1$ is glued to the left half of $\psi_2$. 
In this convention it will be easier to relate algebraic expressions with world-sheet pictures,  
while it is different from the  traditional convention which is widely used in the literatures including Witten \cite{Witten}, Schnabl \cite{Schnabl} or Erler and Maccaferri \cite{ErlerMaccaferri}.  } 
\\

\begin{figure}[tp]
\begin{center}
\begin{overpic}[width=9cm, bb=0 0 1024 768
]{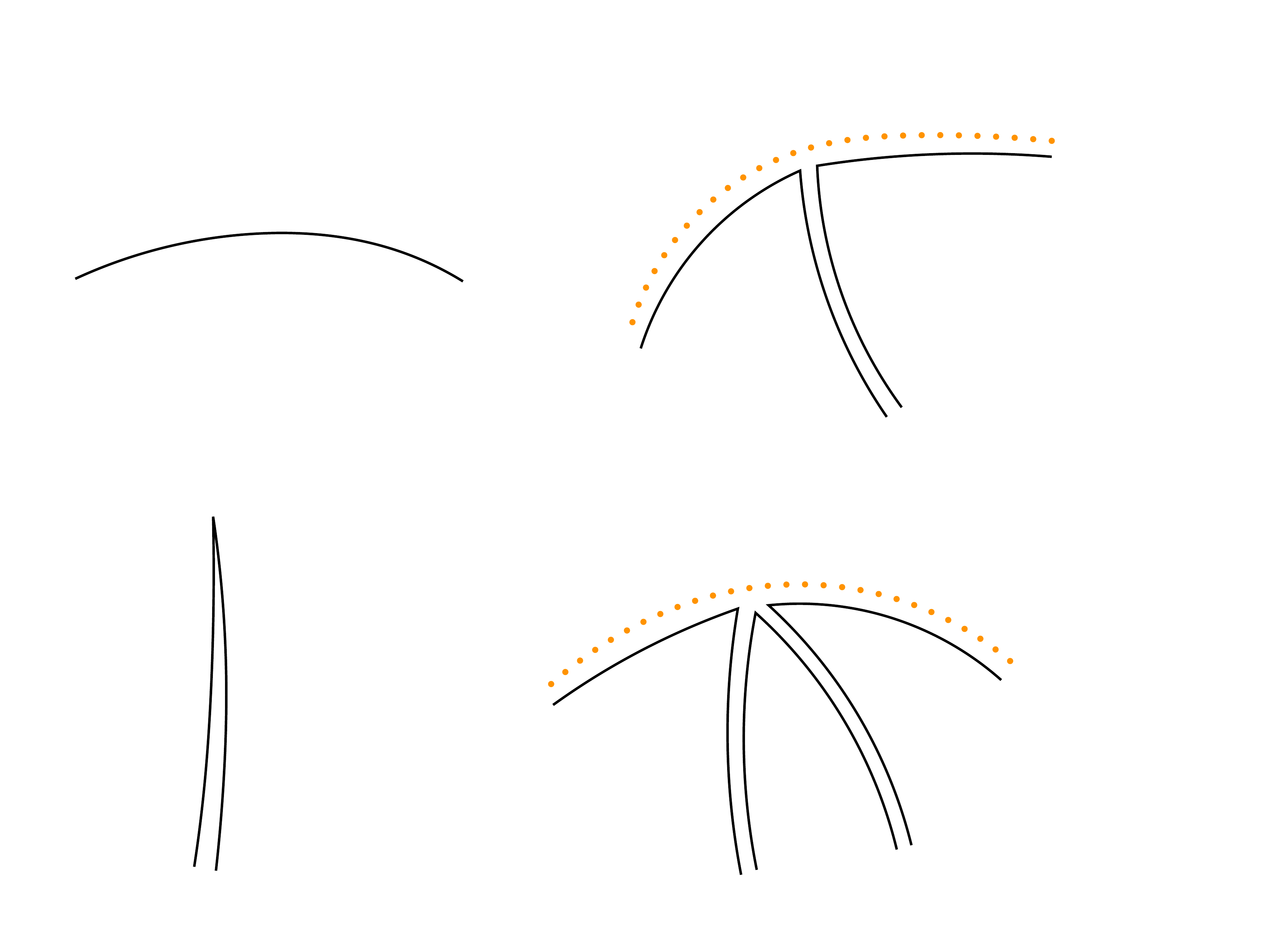}
\put(5,66){\scriptsize(a)}
\put(40,66){\scriptsize(b)}
\put(5,32){\scriptsize(c)}
\put(40,32){\scriptsize(d)}
\put(19.5,49){\scriptsize$\Phi$}
\put(13,52){\scriptsize$L$}
\put(27,52){\scriptsize$R$}
\put(20,55.4){$\cdot$}
\put(19.4,57.55){\scriptsize$M$}
\put(57,47){\scriptsize$\Phi_1$}
\put(74,52){\scriptsize$\Phi_2$}
\put(56,65){\scriptsize$\Phi_1\ast\Phi_2$}
\put(51,31){\scriptsize$\Phi_1\ast\Phi_2\ast\Psi_3$}
\put(16,34.7){\scriptsize$M$}
\put(13,19){\scriptsize$L$}
\put(19,19){\scriptsize$R$}
\put(49,14){\scriptsize$\Phi_1$}
\put(62,13){\scriptsize$\Phi_2$}
\put(71,18){\scriptsize$\Phi_3$}
\end{overpic}
\caption{(a) Witten marked the midpoint $M$ on an open string. The left and the right half is represented by $L$ and $R$, respectively.  
(b)~$\Phi_1\ast \Phi_2$ glues $R$ of $\Phi_1$ with $L$ of $\Phi_2$; $\Phi_1\ast \Phi_2$ appears on the dotted line.  
(c)~$\int \Phi$ glues $L$ and $R$ of $\Phi$.  
(d) This picture illustrates associativity of $\ast$ intuitively. 
} 
\end{center}
\end{figure}
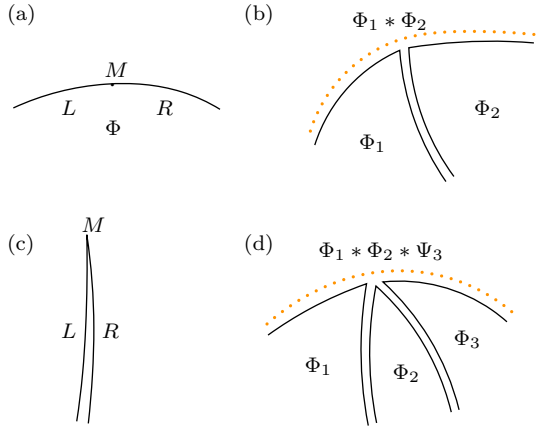

\noindent
For our purpose, it is convenient to rephrase above idea in terms of boundary CFT (BCFT) 
following 
 LeClair, Peskin and Preitshopf (LPP) \cite{LPP1,LPP2}.\footnote{
To be rigorous, the algebraic framework which is derived from LPP's method is slightly different from 
that stated in Section 2.1. It is different  
 in that the star product and the Witten integral are defined only implicitly, and  the existence of the identity string field is not assumed.
There is still controversy as to whether to include the identity string field in $ \mathcal H $.}
This BCFT consists of $c=26$ matter theory and $c=-26$ $bc$-ghost system,
which results from the BRST quantization of open string \cite{Kato:1982im}. 
The vector space $\mathcal H$ is then identified with the state space  of the BCFT,\footnote{
Indeed,  
 we often consider an infinite sum of elements of state space of BCFT, $H_\text{BCFT}$, which is considered to be out of 
$H_\text{BCFT}$. 
A generally accepted definition of $\mathcal H$ is currently lacking.
} $Q$ is the BRST operator, and the degree corresponds to the ghost number of the BCFT.\footnote{
In other words, we consider \textit{the world sheet ghost number}, although  \textit{the spacetime ghost number} and \textit{the total ghost number} are needed in the perturbative calculations  in string field theory. 
In the rest of this paper, the term \textit{ghost number} will be used solely when referring to the world-sheet ghost number.  
  }
Each term of the action \eqref{eq:action} as well as algebraic operations including $\ast$ and $\int$ can all be expressed in terms of BCFT. One such realization will be given in the next subsection.\\

\noindent
Note that the dynamical variable $\Phi$, which is an argument of $S[\Phi]$, should satisfy \textit{the reality condition} so that the action and the physical quantities to be real.  
This is something that must be explained in a proper  review of the open string field theory, 
while we shall not describe it further here. 
See~\cite{Taylor:2003gn} for details. 
\\

\noindent
Of course, we need to choose a boundary condition when we prepare the state space $\mathcal H$. 
In the following, we will refer to this boundary condition by the term \textit{the original D-brane}. 
If it is necessary to specify this boundary condition, the state space will be denoted as $\mathcal H^\text{(b.c.)} $. 
An important working hypothesis is that 
the theory expanded around a classical solution \eqref{eq:tenkaishitasayou} describes an open string theory with another consistent boundary condition, in other words, on another D-brane configuration. 
This expectation naturally stems  from an analogy with usual local field theory. 
A trivial solution $\Psi=0$, which is often referred to as \textit{the perturbative vacuum solution}, describes the original D-brane.  \\

\noindent
In addition, (surprisingly, or some people would say it is natural, but in any case)
there is a classical solution representing the system without D-brane~\cite{Sen:1999xm}. 
We call such a solution a tachyon vacuum solution and denote it by $\Psi_T$.  
In particular, the difference between a classical solution and the tachyon vacuum solution $\Psi' \equiv (\Psi-\Psi_T) $ satisfies
\begin{equation}
\label{eq:eqm_tac}
Q_{T}\Psi'+\Psi'\ast\Psi'=0, 
\end{equation}
where $Q_T\equiv Q_{\Psi_T}$ is the BRST operator around the tachyon vacuum solution. 
The theory expanded around $ \Psi_T $ is expected to have no physical excitation of open strings, which means that  $Q_T$-cohomology at ghost number one is empty.
\\

\noindent
In practice, several types of tachyon vacuum solutions have been obtained in both numerical or analytic ways so far. It is natural to assume that they are all gauge equivalent to each others, though it has not been proved. 
In this paper, we will use the symbol $ \Psi_T $ to represent any of them, for the sake of simplicity.\\

\noindent
Note that 1 will be abused to represent both $1 \in \mathbb C$
 and the identity element of the star product $1\in \mathcal H$, which is often called \textit{the identity string field}. We will also use the notation $|\text{id}\rangle$ to represent the identity string field when we wish to 
 emphasize that it is a state of BCFT.

\subsection{Tools for analytic study}

\noindent
Definition of our new gauge invariant quantities will not need any special tools, and algebraic relations presented in Section 2.1 will suffice.\footnote{To be rigorous, some additional data from BCFT will be necessary to prove the gauge invariance of these quantities completely. 
Furthermore, 
this BCFT data seems to determine whether our formula trivially vanishes. 
See Sections~\ref{sec:manifestly}, \ref{sec:formal} and \ref{sec:huhuihi}. 
}  
To calculate these quantities explicitly, however, some analytic tools obtained in the recent study of classical solutions will be useful. 
Such tools include 
a convenient coordinate system, 
the surface state and the wedge state \cite{Rastelli:2000iu, Schnabl,Schnabl:2002gg} and useful building blocks of classical solutions called $K$, $B$ and $c$~\cite{Okawa}.  \\

\noindent
By the standard state-operator correspondence, 
we see a local operator $\phi(\xi)$
placed at the origin of  the upper half plane (UHP)
defines an open string state $\phi^\text{state}$$ \in \mathcal H$ on the unit semicircle.  
This state $\phi^\text{state}$ can be written as
\begin{equation}
\phi^\textrm{state}=\phi(0) |0\rangle_\text{b.c.},  
\end{equation}
where $|0\rangle_\text{b.c.}$ denotes the conformal vacuum of the BCFT with the boundary condition 
specified by the subscript ($\text{b.c.}$). 
Following Schnabl~\cite{Schnabl},
it is convenient to map this local coordinate to that on $C_2$, which we call \textit{the sliver coordinate}, by (see Figure 2 (a)) 
\begin{equation}
\label{eq:lccccc}
z=f_s(\xi)=\frac{2}{\pi}\arctan \xi,
\end{equation}
where  $C_L$ is the semi-infinite cylinder\footnote{To be precise, $C_L$ is the one-point compactification of  \eqref{eq:bhufid} and is homeomorphic to a disk. } of circumference $L$ defined by the following identification of the upper half plane,
\begin{equation}
\label{eq:bhufid}
z\sim z+L \quad (L>0),\quad  \text{Im}(z)\ge 0. 
\end{equation}
We  will henceforth take a local coordinate $z$ on $C_L$ as $-\frac{L}{2}\le \text{Re}(z)<\frac{L}{2}$. \\

\noindent
Now, all the quantities appearing in the action can be defined with correlation functions on $C_n$:
\begin{equation}
\label{eq:...1}
\int \phi_1^\textrm{state} \ast  \phi_2^\textrm{state}=
\left\langle f_{s}^{(-\frac{1}{2})}\circ \phi_1(0)\,  f_{s}^{(\frac{1}{2})}\circ \phi_2(0)\right\rangle_{C_2}
\end{equation}
\begin{equation}
\label{eq:...2}
\int \phi_1^\textrm{state}  \ast \phi_2^\textrm{state} \ast \phi_3^\textrm{state} =
\left\langle f_{s}^{(-{1})}\circ \phi_1(0)\,  f_{s}^{}\circ \phi_2(0)\,  f_{s}^{(1)}\circ \phi_3(0)\right\rangle_{C_3}
\end{equation}
where $f_{s}^{(x)}(\xi)=f_s(\xi)+x$. 
In addition, \eqref{eq:...1} satisfies the axiom of indefinite inner product (called the BPZ inner product), and \eqref{eq:...2} gives definition of the star product implicitly. \\

\noindent
It might be useful to introduce an intuitive pictorial description of $\ast$ and $\int$, though there is no space to give a full ground for it.  With the theory of surface states in mind, the star product of states is described by an array of strip regions with operator insertions as Figure 2 (c) and (d); the Witten integral is an operation of glueing the left and the right halves of a given state, and then calculating the correlation function on the resulting semi-infinite cylinder (Figure 2 (b)). See also the formula \eqref{eq:ouhukygvhdx}.
\\

\begin{figure}[tp]
\begin{center}
\begin{overpic}[width=13.cm, bb=0 0 1024 768
]{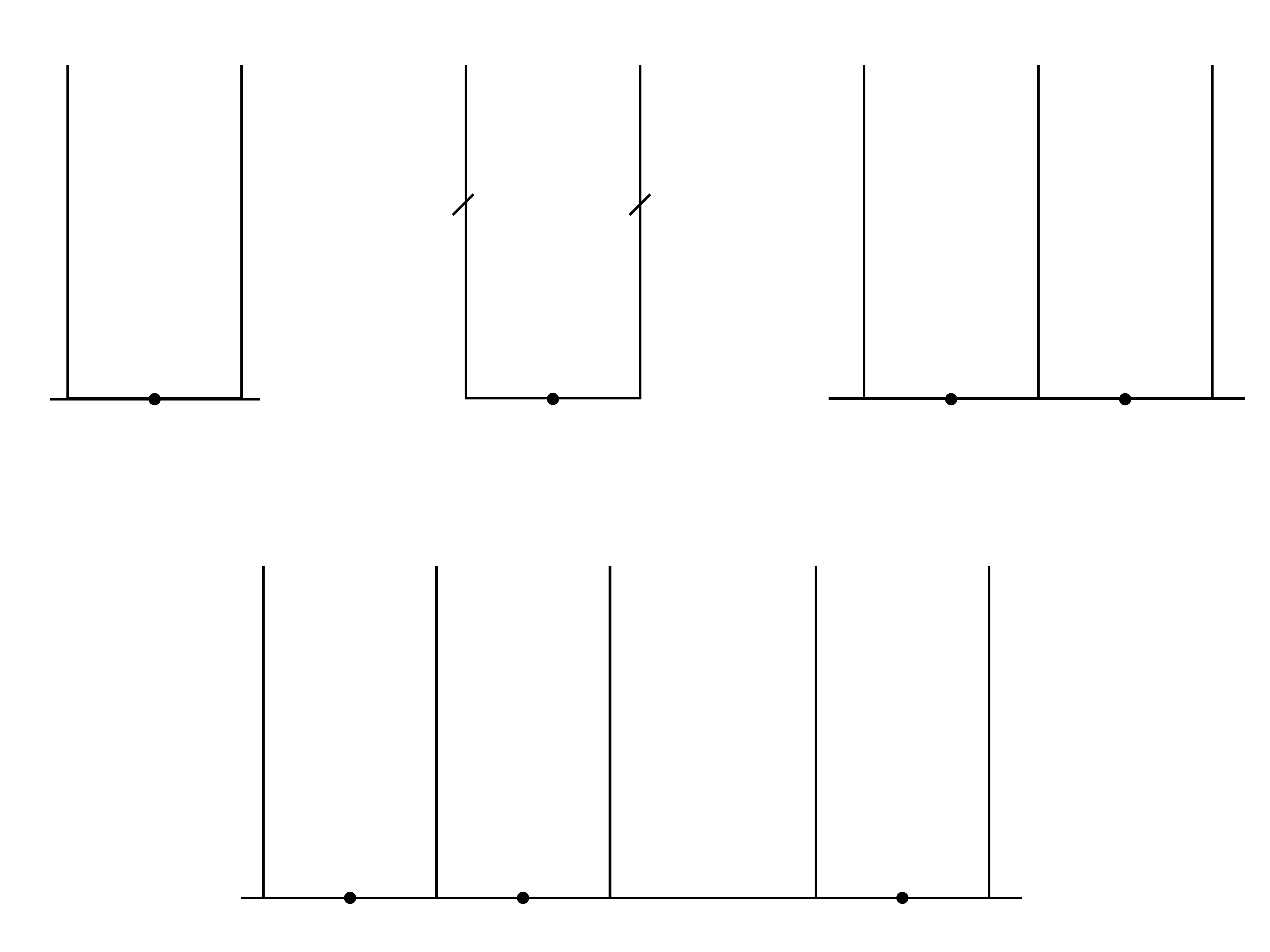}
\put(0,70){\scriptsize(a)}
\put(32,70){\scriptsize(b)}
\put(63.5,70){\scriptsize(c)}
\put(14,30){\scriptsize(d)}
\put(2,40){\scriptsize$-\frac{1}{2}$}
\put(11.5,40){\scriptsize$0$}
\put(72,40){\scriptsize$-\frac{1}{2}$}
\put(87.5,40){\scriptsize$\frac{1}{2}$}
\put(65,40){\scriptsize$-1$}
\put(81,40){\scriptsize$0$}
\put(95,40){\scriptsize$1$}
\put(18,40){\scriptsize$\frac{1}{2}$}
\put(18,1){\scriptsize$-\frac{n}{2}$}
\put(31,1){\scriptsize$-\frac{n-2}{2}$}
\put(44.5,1){\scriptsize$-\frac{n-4}{2}$}
\put(62,1){\scriptsize$\frac{n-2}{2}$}
\put(77.5,1){\scriptsize$\frac{n}{2}$}
\put(8,46){\scriptsize$f_s\circ \phi(0)$}
\put(39.5,46){\scriptsize$f_s\circ \phi(0)$}
\put(68.5,46){\scriptsize$f_s^{(-\frac{1}{2})}\circ \phi(0)$}
\put(84,46){\scriptsize$f_s^{(\frac{1}{2})}\circ \chi(0)$}
\put(23,5.8){\scriptsize$f_1\circ \phi_1(0)$}
\put(37,5.8){\scriptsize$f_2\circ \phi_2(0)$}
\put(66.5,5.8){\scriptsize$f_n\circ \phi_n(0)$}
\put(55,5.8){\scriptsize$...$}
\put(3,59){\scriptsize$L$}
\put(20,59){\scriptsize$R$}
\put(65,59){\scriptsize$L$}
\put(96.2,59){\scriptsize$R$}
\put(17.5,18){\scriptsize$L$}
\put(79,18){\scriptsize$R$}

\end{overpic}
\caption{
(a)  a state $\phi^\text{state}
$ expressed with the local coordinate~\eqref{eq:lccccc} on $C_2$. The unit semi-circle $|\xi|=1$ on UHP is mapped to $z=\pm \frac{1}{2}$.
(b)~$\int \phi$ is given by a correlation function on $C_1$ obtained by identifying $L$ and $R$ of $\phi^\text{state}$. 
(c)~An illustration of $\phi^\text{state}\ast\chi^\text{state}$. 
(d)~An illustration of $\phi_1^\text{state}\ast...\ast \phi_n^\text{state}$, where $f_i(z)=f_s^{(\frac{2i-n-1}{2})}(z)$. 
The Witten integral $\int \phi_1\ast...\ast \phi_n$ is given by a correlation function on $C_n$ obtained by identification of $L$ and $R$ of $\phi_1^\text{state}\ast...\ast \phi_n^\text{state}$. 
}
\end{center}
\end{figure}
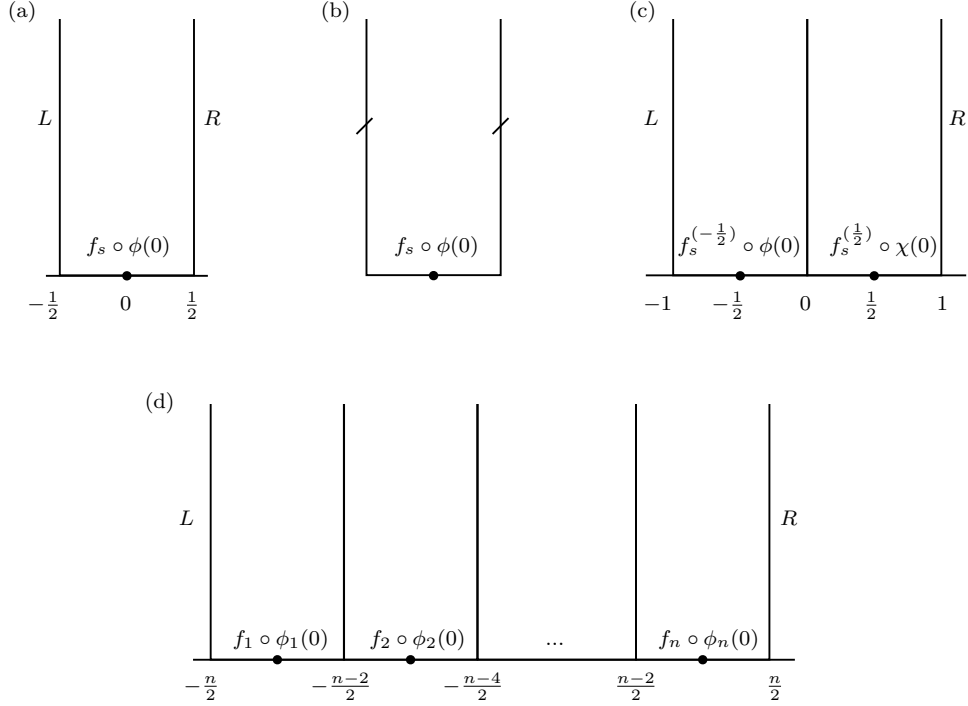

\noindent
Three basic building blocks $K$, $B$, $c$, which are used for construction of analytic solutions, are defined by acting (a line integral of) local operators on the  identity string field:\footnote{Our definition of $\{K, B, c\}$ is different from literatures \cite{Witten, Schnabl, ErlerMaccaferri} because of the convention of the star product  
(see Footnote \ref{foot:leftright}). Fortunately, the rule of translation from one convention to the other is simple. See appendix A of \cite{Erler:2009uj} or Section 2 of  \cite{Okawa}. }
\begin{equation}
K=\int_{i\infty}^{-i\infty} dz T(z)|\text{id}\rangle,\quad B=\int_{i\infty}^{-i\infty} dz b(z)|\text{id}\rangle,\quad c=c(\tfrac{1}{2})|\text{id}\rangle, 
\end{equation}
where we have used the doubling trick to define $T(z)$ and $b(z)$ for Im$(z)<0$.  
They satisfy a series of relations which is often called the $KBc$ algebra
\begin{equation}
\label{eq:kbc}
[K, B]=0,\quad [B,c]=1, \quad Qc=cKc, \quad QB=K.
\end{equation}
Note that the graded commutator $[\phi_1,\phi_2]$ is defined with the star product by $\phi_1\ast \phi_2-(-)^{|\phi_1||\phi_2|}\phi_2\ast \phi_1$. 
In this paper, we often omit the symbol $\ast$ , especially when we work with $K,B,c$, 
as in the third equation of \eqref{eq:kbc}. 
We also define $O\in \mathcal H$ corresponding to a general local boundary operator $O(x)$ in a similar fashion as $c$,  
\begin{equation}
\label{eq:local_ins_identity}
O\equiv  O(\tfrac{1}{2})|\text{id}\rangle.
\end{equation}

\noindent
Another basic tool we would like to introduce  is the wedge state $|r\rangle$. For $r\in \mathbb N$, it is simply given by
\begin{equation}
|r\rangle=
\overbrace{
 |0\rangle_\text{b.c.} 
\ast
...
\ast 
 |0\rangle_\text{b.c.}}^{r}. 
\end{equation}
This definition can naturally be extended to a non-negative real number $r$. 
\noindent
By using $K$, the wedge state can be characterized by the following differential equation
\begin{equation}
\frac{d}{dr}|r\rangle=K\ast |r\rangle
\end{equation}
with the initial condition $|r=1\rangle=|0\rangle_\text{b.c.}$. It is thus natural to write 
\begin{equation}
e^{rK} =|r\rangle, \quad r\ge0.
\end{equation}
The wedge state is related to the correlation function on a semi-infinite cylinder by the following relation
\begin{equation}
\label{eq:ouhukygvhdx}
\begin{split}
&\int e^{x_1 K} \ast O_1 \ast e^{x_2 K} \ast O_2 \ast ...\ast e^{x_n K} \ast O_n \ast e^{x_{n+1} K}\\
=&\left\langle O_1(y_1)O_2(y_2)... O_n(y_n)\right\rangle_{C_L},
\end{split}
\end{equation}
with
\begin{equation}
y_i=x_1+...+x_{i}-\frac{L}{2},
\quad  L=x_1+...+x_{n+1}. 
\end{equation}

\subsubsection*{Correlation functions on a semi-infinite cylinder}

\noindent
Here we exhibit a few formulas for basic correlation functions on  $C_L$.  For the ghost part,  we introduce
\begin{equation}
f_{123}\equiv
\big\langle c(z_1) c(z_2) c(z_3)
\big\rangle_{C_L}^\text{ghost}=\frac{L^3}{\pi^3}
\sin\frac{\pi z_{21}}{L}
\sin\frac{\pi z_{32}}{L}
\sin\frac{\pi z_{31}}{L}
\end{equation}
and
\begin{equation}
\label{eq:4ptghostcrrtr}
\big\langle Bc(z_1) c(z_2) c(z_3)c(z_4)\big\rangle_{C_L}^\text{ghost}=\frac{1}{L}(-{z_1}f_{234}+{z_2}f_{134}-{z_3}f_{124}+{z_4}f_{123}).
\end{equation}
For the matter part, we will use the four point function of  on-shell tachyon states 
with the Neumann boundary condition
\begin{equation}
\begin{split}
&\left\langle T_1(z_1) T_2(z_2) T_3(z_3) T_4(z_4) \right\rangle_{C_L}^\text{matter, Neumann}\\
&=
iC\frac{\pi^4}{L^4}\prod_{i<j}  \left|  \sin\frac{\pi z_{ij}}{L}\right|^{2\alpha'k_i\cdot k_j} \delta\left(\sum_i k_i \right), 
\end{split}
\end{equation}
where $C$ is a normalization constant, and
$T_j(z)$ represents the vertex operator of on-shell tachyon 
\begin{equation}
\label{eq:ttt}
 T_j(z)= ce^{i k_j \cdot X}(z), \quad
(k_j)^2= \frac{1}{\alpha'}.
\end{equation}

\subsubsection*{Ellwood correspondence}

\noindent
Finally, we will briefly describe Ellwood's interpretation of \textit{the gauge invariant observable} which we mentioned in Section 1. %
The gauge invariant observable referred to here is the following quantity, which is defined against a classical solution $ \Psi $ and an on-shell closed string vertex operator $ V(z)=c\bar c\mathcal V (z) $
\begin{equation}
\int V(i\infty )\Psi,
\end{equation}
which means that we insert $V(z)$ at the midpoint of $\Psi$ ($z=i\infty$ in the sliver coordinate \eqref{eq:lccccc}) and then perform the Witten integral. 
This quantity is often called \textit{the Ellwood invariant} after his study \cite{Ellwood}, while it was originally discovered by Hashimoto-Itzhaki \cite{HI1} or other authors  \cite{GRSZ1, Z1}. 
Following \cite{Kudrna:2012re}, it is convenient to rewrite it by using a reference tachyon vacuum solution $\Psi_T$
\begin{equation}
\label{eq:kjhgfs}
\int \mathcal V(i\infty )(\Psi-\Psi_T)\equiv I_\text{E}(\Psi).
\end{equation}
Ellwood conjectured that $I_\text{E}(\Psi)$ is equal to the tadpole amplitude of an on-shell closed string on the D-brane configuration represented by the classical solution $\Psi$.  
For a partial proof of this conjecture, see Baba and Ishibashi~\cite{Baba:2012cs}.


\noindent

\section{Gauge invariant quantity for three external states}

\noindent
Now, let us start our discussion on the new gauge invariant quantity for three external states. 
In Section 3.1, we will consider  the quantity around the trivial classical solution $ \Psi = 0 $. 
Some important properties such as decoupling of the null states will be shown. 
In Section 3.2, we will extend the definition of this quantity to the general classical solutions and will discuss its invariance under the gauge transformation of the classical solution.

\subsection{... around the perturbative vacuum}
As we saw in \eqref{eq:kjhgfs}, 
it would be convenient for our purpose to introduce a tachyon vacuum solution $\Psi_T $ as a reference.  
The trivial solution $\Psi=0$ can be written as 
\begin{equation}
0= \underbrace{\Psi_T}_{\text{reference}} \quad +\quad (-\Psi_T),
\end{equation}
 where the second term on the right hand side
 is a solution to \eqref{eq:eqm_tac}. 
This $(-\Psi_T)$ is interpreted as 
the original D-brane, and it would be natural to use $(-\Psi_T)$ 
 to write down a physical quantity for the perturbative vacuum. \\

\noindent
Another natural object defined around $\Psi_T$ would be \textit{a homotopy state}
satisfying
\begin{equation}
\label{eq:htp}
Q_T A_T=1, 
\end{equation}
which is used to show that  the cohomology of $Q_T$ is empty\footnote{
For most of known tachyon vacuum solutions \cite{Schnabl,Erler:2009uj,Sen-Zwiebach,Takahashi:2002ez}, 
an expression for $A_T$ is known \cite{Ellwood+Schnabl,EFHM,Inatomi:2011xr,Inatomi:2011an}. 
Note that 
there are some numerical results which support 
$Q_T$-cohomology at some non-standard ghost numbers is not empty \cite{GI,I}.
}.
From these two objects, we define $W$ by\footnote{
It is interesting that the combination \eqref{eq:WWWWW} has already appeared in Ellwood's another seminal paper \cite{Ellwood:2009zf} to define his \textit{characteristic projector}. 
The authors would like to thank C. Maccaferri for drawing their attention to this fact. 
}
\begin{equation}
\label{eq:WWWWW}
  W=A_T  (-\Psi_T)+ (-\Psi_T)A_T.
\end{equation}
This quantity is $Q$-closed:
\begin{equation}
  QW=0.
\end{equation}
As we will see in the later sections,  $W$ and $A_T$ will play a fundamental role in our construction of the gauge invariant quantity. \\
\vspace{5mm}

\noindent
As an example, let us calculate  $W$  for a simplest case and see what it looks like. 
Consider the Erler-Schnabl solution \cite{Erler:2009uj} as the reference $\Psi_T$:
\begin{equation}
\label{eq:ESs}
\Psi_T=-\frac{1}{\sqrt{1-K}}c(1-K)Bc\frac{1}{\sqrt{1-K}}.
\end{equation}
We can take the following $A_T$
\begin{equation}
\label{eq:Akantan}
  A_T=-\frac{B}{1-K}\quad\left(\equiv\int_0^\infty dxe^{-x}Be^{xK}\right), 
\end{equation}
and  $W$ is given by
\begin{equation}
\label{eq:Wkantan}
  W=-\frac{1}{1-K} \quad\left(\equiv\int_0^\infty dxe^{-x}e^{xK}\right).
\end{equation}
As we see, $A_T$ and $W$ are superposition of wedge states with a line integral of $b$-ghost  for $A_T$,   
and they look like a fragment of open string propagator or worldsheet, respectively.\footnote{
For tachyon vacuum solutions of the form \cite{Okawa}
\begin{equation}
\Psi_T=Fc\frac{KB}{1-F^2}cF
\end{equation}
with $F=F(K)$, we can take 
\begin{equation}
A_T=\frac{1-F^2}{K}B
\end{equation}
and $W$ is then simply given by
\begin{equation}
W=-F^2. 
\end{equation}
} 
\\

\noindent
Let us then try expressing an on-shell three-point amplitude  of tachyons using $W$ in~\eqref{eq:Wkantan} and  
 $T_i\in \mathcal H$ defined by \eqref{eq:ttt} with \eqref{eq:local_ins_identity}: 
\begin{equation}
\label{eq:3ptexp}
\begin{split}
  &\int W T_1 W T_2 W T_3\\
  =&\prod_{j}^3\int_0^\infty dx_j e^{-x_j}\left \langle ce^{ik_1 \cdot X}(y_1) ce^{ik_2 \cdot X}(y_2) ce^{ik_3 \cdot X}(y_3) \right\rangle_{C_{L}}\\
 \propto& i\delta(k_1+k_2+k_3),
  \end{split}
\end{equation}
where 
\begin{equation}
y_i=\sum_{j=1}^i x_j-\frac{x_1+x_2+x_3}{2}.
\end{equation} 
This might seem like a trivial rephrasing of the known result; however, we would like to claim that the formula (\ref{eq:3ptexp}) has essential meaning. 
Indeed, the agreement with the scattering amplitude is true regardless of the choice of~$\Psi_T$.
\\

\noindent
Let us 
return to discussion with general $\Psi_T$. 
Consider the following quantity, which is a generalization of \eqref{eq:3ptexp},
\begin{equation}
\label{eq:3ptgen1}
 I_0= \int W \mathcal O_1 W \mathcal O_2 W \mathcal O_3,
\end{equation}
where $\mathcal O_j$ is a $Q$-closed state 
\begin{equation}
 Q \mathcal O_i=0.
\end{equation}

\noindent
This $I_0 $ satisfies the following properties:
\begin{enumerate}
  \item[(a)] $I_0 $ is invariant under replacing the reference tachyon vacuum solution
  \begin{equation}
  \label{eq:3pt_transf_a}
  \begin{cases}
  \Psi_T\mapsto \Psi_T'=U^{-1}(\Psi_T+Q)U \\
  A_T\mapsto U^{-1}A_TU.
  \end{cases}
\end{equation}

   \item[(b)]  $I_0 $ is independent of the choice of $A_T$. In other words, it is invariant under the following transformation:
   \begin{equation}
   \label{eq:3pt_transf_b}  
  A_T\mapsto A_T+Q_T(...). 
\end{equation}

    \item[(c)]  $I_0 $ is invariant under the linearized gauge transformation of $\mathcal O_i$
    \begin{equation}
    \label{eq:iiuhiuh}
  \mathcal O_i \mapsto \mathcal O_i+Q(,,,).
\end{equation}

\end{enumerate}

\noindent
\textbf{Proof}\quad 
Since both $W$ and $\mathcal O_j$ are $Q$-closed, $I_0$ does not change if we change 
$W$ or $\mathcal O_j$ by $Q$-exact terms. 
That verifies the invariance under \eqref{eq:iiuhiuh}. 
Under the transformation \eqref{eq:3pt_transf_a}, $W$ changes as
\begin{equation}
W \mapsto W +Q Y
\end{equation}
with
\begin{equation}
Y=U^{-1}A_T U - A_T. 
\end{equation}
Finally, under the transformation \eqref{eq:3pt_transf_b}, $W$ changes as 
\begin{equation}
W\mapsto W+QQ_T(...). 
\end{equation}
We therefore conclude that  $I_0$ is invariant under \eqref{eq:3pt_transf_a}, \eqref{eq:3pt_transf_b} and \eqref{eq:iiuhiuh}. $\square$\\

\noindent
Now, it is evident that $I_0$ gives on-shell three point amplitudes. \\

\subsection{... around generic classical solutions}
We next would like to define the similar quantity around an arbitrary classical solution $\Psi$. As before, the difference from the reference $\Psi_T$ is what is relevant for us
\begin{equation}
\Psi=\underbrace{\Psi_T}_\text{reference}\quad +\quad(\Psi-\Psi_T).
\end{equation}
We then define  $W_\Psi$ by
\begin{equation}
  W_\Psi=A_T(\Psi-\Psi_T)+ (\Psi-\Psi_T) A_T.
\end{equation}
This $W_\Psi$ is $Q_\Psi$-closed:
\begin{equation}
  Q_\Psi W_\Psi=0.
\end{equation}
Now, we can define the gauge invariant quantity $I_\Psi$ using $W_\Psi$ as 
\begin{equation}
  I_\Psi=\int \mathcal O_i W_\Psi\mathcal O_j W_\Psi\mathcal O_k W_\Psi,
\end{equation}
where $ \mathcal O_i \in \mathcal H$ satisfies  the following conditions:
\begin{equation}
\label{eq: psc1}
Q_\Psi \mathcal O_i=0,
\end{equation}
\begin{equation}
\label{eq: psc2}
  |\mathcal O_i|= 1.
\end{equation}
The equation \eqref{eq: psc1} is considered as the physical state condition around the classical solution $\Psi$ naturally (see  \cite{Hene?,KO1,KO2} for general discussion). Throughout this paper, the term \textit{an external state} refers to a state $\mathcal O_i$ which satisfies  \eqref{eq: psc1} and \eqref{eq: psc2}. 
\\ 

\noindent
In the rest of this subsection, we will show various algebraic properties of~$I_\Psi$.

\begin{enumerate}
  \item[\{1\}] $I_\Psi$ is invariant under replacing the reference tachyon vacuum solution
  \begin{equation}
   \label{eq:trnsf1}
  \begin{cases}
  \Psi_T\mapsto \Psi_T'=U^{-1}(\Psi_T+Q)U \\
  A_T\mapsto U^{-1}A_TU
  \end{cases}
\end{equation}

 \item[\{2\}] $I_\Psi $ is invariant under gauge transformation of the classical solution
  \begin{equation}
  \label{eq:trnsf2}
  \begin{cases}
  \Psi\mapsto \Psi'=U^{-1}(\Psi+Q)U \\
  \mathcal O_i \mapsto U^{-1}\mathcal O_iU
  \end{cases}
\end{equation}

   \item[\{3\}]  $I_\Psi $ is independent of the choice of $A_T$; that is, $I_\Psi$ is invariant under the following transformation:
   \begin{equation}
   \label{eq:trnsf3}
  A_T\mapsto A_T+Q_T(...)
\end{equation}

    \item[\{4\}]  $I_\Psi$ depends only on a $Q_\Psi$-cohomology class of $\mathcal O_j$. 
     In other words, $I_\Psi $ is invariant under the following transformation:
    \begin{equation}
    \label{eq:trnsf4}
  \mathcal O_i \mapsto \mathcal O_i+Q_\Psi(,,,).
\end{equation}

\end{enumerate}

\noindent
\textbf{Proof}
\quad 
Since both $W_\Psi$ and $\mathcal O_j$ are $Q_\Psi$-closed, $I_\Psi$ does not change if we change $W_\Psi$ or $\mathcal O_j$ by $Q_\Psi$-exact terms. Under the transformation \eqref{eq:trnsf1}, $W_\Psi$ changes as
\begin{equation}
W_\Psi \mapsto W'_\Psi =W_\Psi +Q_\Psi Y
\end{equation}
with
\begin{equation}
Y=U^{-1}A_T U - A_T. 
\end{equation}
Under the transformation \eqref{eq:trnsf3}, $W_\Psi$ changes as 
\begin{equation}
W_\Psi\mapsto W_\Psi+Q_\Psi Q_T(...). 
\end{equation}
Therefore, $I_\Psi$ is invariant under \eqref{eq:trnsf1}, \eqref{eq:trnsf3} and \eqref{eq:trnsf4}. 
To prove the invariance under \eqref{eq:trnsf2}, notice that the combination of \eqref{eq:trnsf1} and  \eqref{eq:trnsf2} 
is a similarity transformation%
  \begin{equation}
  \begin{cases}
  \label{eq:all_sim_transf}
  (\Psi-\Psi_T)&\mapsto U^{-1}( \Psi-\Psi_T)U \\
  A_T&\mapsto U^{-1}A_TU\\
    \mathcal O_i &\mapsto U^{-1}\mathcal O_iU, 
  \end{cases}
\end{equation} 
which does not change $I_\Psi$. We therefore conclude that $I_\Psi$ is also invariant under~\eqref{eq:trnsf2}. 
$\square$\\

\section{Gauge invariant quantity for $N$ external states}

\noindent
In this section, we will give a definition of the gauge invariant quantity $I^{(N)}_\Psi$ for $N$ external states. 
Sections 4.1 and 4.2 will focus on the case of $N=4$.
In Section~4.1 we will define a quantity $H_\Psi$, which is gauge invariant under some additional condition~\eqref{eq:simplef2} on the external states. In Section 4.2 we present a definition of the boundary term $K_\Psi$, and the sum $I^{(4)}_\Psi =H_\Psi+K_\Psi$ will be gauge invariant without any additional condition. 
Section 4.3 will focus on the case of $N\ge 4$. \\

\subsection{Simple formula}
In order to define four point gauge invariant quantities, we need to neutralize the extra ghost number coming from the external states. 
The following expression would be a natural guess for such a formula:
\begin{equation}
\label{eq:simplef}
  H_\Psi=\frac{1}{4}\sum_{\sigma}  \int H_{\sigma(1)\sigma(2)\sigma(3)\sigma(4)}
\end{equation}
with
\begin{equation}
\label{eq:simplef2}
  H_{ijkl}=  \int A_T \mathcal O_{i} W_\Psi \mathcal O_{j} W_\Psi\mathcal O_{k} W_\Psi\mathcal O_{l},
\end{equation}
where $\sigma$  runs over the permutations $\sigma \in S_4$ of the indices $\{1,2,3,4\}$. 
This $H_\Psi$ is \textit{not} gauge invariant as it is; however, when the external states $\{\mathcal O_i\}$ satisfy the following orthogonality condition 
\begin{equation}
\label{eq:orthogonality}
  \mathcal O_i \mathcal O_j=0,  \quad i\ne j,
\end{equation}
$H_\Psi$ indeed defines a gauge invariant quantity.\\

\noindent
In Section 5, we will calculate $H_\Psi$ corresponding to a four point amplitude of on-shell tachyon states. It will be shown that the condition \eqref{eq:orthogonality} will be satisfied by imposing appropriate conditions on the external momentum.

\subsubsection*{Proof of invariance under the transformations \eqref{eq:trnsf1} - \eqref{eq:trnsf4}}

\noindent
Let us first prove invariance under replacement of the tachyon vacuum solution \eqref{eq:trnsf1}. 
Under this transformation, 
$A_T$ and $W_\Psi $ transform as
\begin{equation}
\label{eq:A1}
A_T\to A_T'=A_T+Y,
\end{equation}
\begin{equation}
\label{eq:W1}
W_\Psi \to W_\Psi '=W_\Psi +Q_\Psi Y, 
\end{equation}
\noindent
where $Y=UA_TU^{-1}-A_T$. 
The transformation of $H_{1234}$ is 
\begin{equation}
\label{eq:transfH1234}
\begin{split}
H_{1234}\mapsto H_{1234}&+T_{[]1}-T_{[]2}+T_{[]3}-T_{[]4}\\
                          &+T_{[2]1}+T_{[3]1}+T_{[4]1}-T_{[3]2}-T_{[4]2}+T_{[4]3}\\
                          &+T_{[23]1}+T_{[24]1}+T_{[34]1}-T_{[34]2}\\
                          &+T_{[234]1}\\
                          &+(\text{contact terms})
\end{split}
\end{equation}
with $T_{[i_1...i_m]j}$ defined by the following rule: consider an expression
\begin{equation}
\int Z\mathcal O_1 Z\mathcal O_2Z\mathcal O_3 Z\mathcal O_4.
\end{equation}
We replace all the $i_1,...,i_m$-th $Z$ by $(Q_\Psi Y)$; $j$-th $Z$ by $Y$; all the other $Z$'s by $W_\Psi $. The outcome is what $T_{[i_1...i_m]j}$ denotes. 
This $T_{[i_1...i_m]j}$ is totally symmetric under the exchange of the indices inside the 
square brackets. 
 It also has the property
\begin{equation}
T_{[i_1...i_{m-1}x]y}=T_{[i_1...i_{m-1}y]x}. 
\end{equation}
The term (contact terms) in \eqref{eq:transfH1234} represents a set of terms in which 
a near collision of the two external states
$\mathcal O_i\mathcal O_j$ occurs inside the Witten integral. 
We can drop this part using the orthogonality condition \eqref{eq:orthogonality}.\\

\noindent
The remaining part (which is sum of $T_{[i_1...i_m]j}$'s) changes its sign under the cyclic permutation of the indices $(1,2,3,4)\to (4,1,2,3)$. This means the total $H_\Psi$ is invariant under \eqref{eq:trnsf1}. \\

\noindent
Note that the invariance of $H_\Psi$ under the replacement of $A_T$ \eqref{eq:trnsf3} also follows from the above discussion, 
because the change of $A_T$ and $W_\Psi$ under \eqref{eq:trnsf3} take the same form as \eqref{eq:A1} and \eqref{eq:W1}. 
Since $H_\Psi$ is invariant under the similarity transformation \eqref{eq:all_sim_transf}, we can also conclude that $H_\Psi$ is invariant under the gauge transformation \eqref{eq:trnsf1}. 
Proving the invariance under the transformation~\eqref{eq:trnsf4} is 
very similar to before except for to deal with $A_T$. 
$\square$\\


\subsection{Complete formula}
\label{sec:manifestly}

It is possible to write down a manifestly gauge invariant formula for 
$N= 4$ without an artificial condition \eqref{eq:orthogonality}. 
To do this, however, we need to introduce an apparently formal object, namely a state $A_\Psi$ which satisfies
\begin{equation}
\label{eq:mitasenai}
  Q_\Psi A_\Psi=1.
\end{equation}

\noindent
By using $A_\Psi$, we can define the gauge invariant quantity $I_\Psi^{(4)}$ for four external states as
\begin{equation}
\label{eq:4pt}
I_\Psi^{(4)}=H_\Psi +K_\Psi,
\end{equation}
where
\begin{equation}
\label{eq:surface}
K_\Psi=-\frac{1}{4}\sum_{\sigma}  \int A_\Psi \mathcal O_{\sigma(1)} W_\Psi \mathcal O_{\sigma(2)} W_\Psi\mathcal O_{\sigma(3)} W_\Psi\mathcal O_{\sigma(4)}.
\end{equation}
Since the cohomology of $Q_\Psi$  is not empty in general, there is no $A_\Psi$ which satisfies \eqref{eq:mitasenai} in any definite sense; however,  this quantity $A_\Psi$ can be defined when it is placed at a proper place in a correlation function. In this sense, the formula \eqref{eq:4pt} is defined as a calculable quantity. 
This might be compared to the process of defining 
a propagator as 
the inverse of a differential operator
 in quantum field theory.\\

\noindent
The new term $K_\Psi$, and also the contributions from it, will be referred to as \textit{the boundary term}. 
The role of this term closely resembles the boundary term presented in a recent paper by Sen \cite{Sen:2019jpm}, 
as we will see in Section 5. 
\\

\noindent
Proof of the invariance of 
$I_\Psi^{(4)}$
under a replacement of 
the reference $\Psi_T$ 
is the same as before, except for absence of the contact term. 
In the present case, no contact terms appear because they are canceled by the boundary term. \\

\noindent
If we look closer, we would 
find
 a 
latent defect 
 in this proof; that is, whether we can discard the surface term of the Witten integral $\int Q_\Psi(...)$ is not clear. 
This $(...)$ part has some factors of $A_\Psi$, which is not necessarily an element of $\mathcal H$ (typically, $\frac{B}{K}$ appears in $(...)$).
Let us further study this point with a concrete example in Section \ref{sec:huhuihi}, where we will see that this surface term 
vanishes in spite of this delicateness.  
The proof of gauge invariance of $I_\Psi^{(4)}$ will also be completed there.

\subsection{Extension to $N$ point invariants}
It is straightforward to generalize the formula \eqref{eq:4pt} for $N$ external states:
\begin{equation}
\label{eq:nten}
I_\Psi^{(N)}=C_N\sum_{\sigma}\int \prod_{i=1}^N \left[ (W_\Psi+A_T-A_\Psi) \mathcal O_{\sigma(i)}\right], 
\end{equation}
where $C_N$ is an overall combinatorial constant depending only on $N$.   \\

\noindent
Let us prove the invariance of $I_\Psi^{(N)}$ under the infinitesimal version of the transformation \eqref{eq:trnsf1}. 
To start with, rewrite $I_\Psi^{(4)}$ as
\begin{equation}
I_\Psi^{(4)}=C_N\sum_{\sigma }\sum_{(p,q,r)}H^{(p,q,r)}_{\sigma(1)\sigma(2)...\sigma(N)}  
\end{equation}
where $H^{(p,q,r)}_I$ $(I\equiv i_1,...,i_N)$ is defined by the following rule: consider an expression
\begin{equation}
\label{eq:lalalala}
\int Z\mathcal O_{i_1} Z\mathcal O_{i_2}...Z\mathcal O_{i_N}.
\end{equation}
We replace all the $p,q,r$-th $Z$ by $W_\Psi$; all the other $Z$'s by $A=A_T-A_\Psi $. The outcome is what $H_I^{(p,q,r)}$ denotes. \\

\noindent
To systematically represent the variation of $H^{(p,q,r)}_{I}$, let us also define
$T^{j(p,q,r)}_I $  and $T^{[j](p,q)}_I $:
\begin{itemize}
\item $T^{j(p,q,r)}_I$  is defined by  \eqref{eq:lalalala}
with replacing $p,q,r$-th $Z$ by $W_\Psi$; $j$-th $Z$ by $Y$; all the other $Z$'s by $A=A_T-A_\Psi $. 

\item $\tilde T^{[j](p,q)}_I$  is defined by \eqref{eq:lalalala}
with replacing  $p,q$-th $Z$ by $W_\Psi$; $j$-th $Z$ by $Q_\Psi Y$; all the other $Z$'s by $A=A_T-A_\Psi $. 
\end{itemize}

\noindent
With these notations, variation of $H_{I}^{(p,q,r)}$ is written as
\begin{equation}
\label{eq:sikisiki2}
\delta H_{I}^{(p,q,r)}=\sum_{i\notin \{p,q,r\}}
T^{i(p,q,r)}_{I}+\tilde T^{[p](q,r)}_I+\tilde T^{[q](p,r)}_I+\tilde T^{[r](p,q)}_I.
\end{equation}
Note that we regarded $Y$ as an infinitesimal and kept the terms up to the first order of it. From the Leibniz rule of $Q_\Psi$, it follows that
\begin{equation}
\label{eq:sikisiki3}
\tilde T^{[i](p,q)}_J=\sum_{r\ne p,q,i} {s(r,i;p,q)}T^{i(p,q,r)}_J,
\end{equation}
where ${s(r,i;p,q)}\in \{-1,+1\}$ is a sign factor defined by the following rule:
\begin{equation}
s(w,x;y,z)=s(x,w;y,z)=s(w,x;z,y),
\end{equation}
\begin{equation}
 s(w,x;y,z)=-s(w,y;x,z),
\end{equation}
\begin{equation}
 s(w,x;y,z)=-1\quad \text{if }\max(w,x)<\min(y,z). 
\end{equation}
Now, consider the partial sum 
\begin{equation}
\sum_{(x,y,z)}\delta H^{(x,y,z)}_J.
\end{equation}
From \eqref{eq:sikisiki2} and \eqref{eq:sikisiki3}, this sum can be expressed as a linear combination of $T_J^{i(p,q,r)}$. Let us then focus on the coefficient of $T_J^{i(p,q,r)}$. 
The source of $T_J^{i(p,q,r)}$ is fourfold: $\delta H^{(p,q,r)}_J$, $\delta H^{(i,p,q)}_J$, $\delta H^{(i,p,r)}_J$ and $\delta H^{(i,q,r)}_J$.  We can confirm that these four contributions exactly cancel by checking their sign factors.   
We thus find $\delta I_\Psi^{(N)}=0$. \\

\noindent
Let us omit the proof of invariance under the transformations \eqref{eq:trnsf2} - \eqref{eq:trnsf4} because  
the proof is similar as that for $N=4$.

\subsection{Formal expression as a surface term}
\label{sec:formal}

\noindent
As we showed, most of our results can be obtained by using the formal algebraic rules stated in Section 2.1 only; however, it necessitates (analytic) tools of BCFT introduced in Section 2.3 to evaluate $I_\Psi^{(N)}$. 
Without implementation by BCFT, it is also not possible to elucidate some subtle queries on our construction, as we will see in this 
subsection. 
\\

\noindent
For a long period after finding \eqref{eq:nten},  the authors were haunted by a suspicion that our gauge invariant quantity is simply zero.   
Indeed, 
it admits an expression as a formal surface term of the Witten integral
\begin{equation}
\label{eq:bizarre}
I_\Psi^{(N)} \overset{?}{=} \sum_{(k,l)} \int Q_\Psi R_{k,l} \quad \textbf{(FORMAL)},
\end{equation}
where $R_{k,l}$ is algebraically defined by the following product
\begin{equation}
Z \mathcal O_1Z \mathcal O_2...Z \mathcal O_N
\end{equation}
with 
replacing $k$-th and $l$-th $Z$'s by $W_\Psi $, and 
 all the other $Z$'s by $A=A_T-A_\Psi$. 
Of course, we know that such a formal trivialization argument is not uncommon: as claimed in another seminal paper of Ellwood \cite{Ellwood:2009zf}, all the classical solutions can formally be written as the pure gauge form (see also Erler and Maccaferri \cite{Erler:2012qn}). \\

\noindent
In fact, it is not possible to construct $ R_ {k, l} $ as an element of $ \mathcal H $ and then the gauge invariant quantity is not identically zero; however, 
we never know such a thing as far as thinking only with the algebra in Section 2.1.\\

\noindent
The expression \eqref{eq:bizarre} looks more interesting 
if we compare it with the fact stated in Section 4.2. 
We will return to this problem again in Section \ref{sec:huhuihi} and give an explanation on the difference between them.

\section{Physical interpretation}

In this section, we present  calculations of $I_\Psi^{(4)}$ for the case which the classical solution $\Psi$ is the Erler-Maccaferri solution~\cite{ErlerMaccaferri}.  As the title of their paper suggests, the Erler-Maccaferri solution can represent any possible boundary condition of open string theory.\footnote{
The claim of  the original paper \cite{ErlerMaccaferri} was that any   $X^0$-independent 
D-brane configuration can be described by a classical solution of the form \eqref{eq:EMsolution};
however, 
in recent talks \cite{Erler_talk} it is declared that, with a modified construction of ${\Sigma}$ and $\bar{\Sigma}$, we can remove this restriction while keeping the form of the solution \eqref{eq:EMsolution} and algebra \eqref{eq:algebra1} - \eqref{eq:algebra3} untouched. 
} 
We therefore believe that the calculations in this section have generality to some extents.

\subsection{Formula}
Before starting our discussion, let us introduce a convenient formula for the correlation function on a semi-infinite cylinder $C_L$: for $0\le a, b$
\begin{equation}
\label{eq:formula}
\begin{split}
&\int_a^b dx 
\langle BO_1(z_1)  O_2(z_2) O_3(z_3) O_4(z_4) \rangle_{C_L}\\
=
&
\int_{F(a)}^{F(b)} du
\langle  V_1(u)  V_2(1) V_3(\infty) V_4(0) \rangle_{\text{UHP}}^\text{matter, (b.c.)},
\end{split}
\end{equation}
where, the positions of boundary operators are given by $-\frac{L}{2}<z_1<z_2<z_3<z_4<\frac{L}{2}$ with $L=x+(z_4-z_1)$; 
$O_j(z)$ is a $Q$-closed boundary operator 
of the following form
\begin{equation}
\label{eq:ocv}
O_j(z)=c V_j(z),
\end{equation}
with $V_j(z)$ the matter part of $O_j(z)$;
(b.c.) represents some boundary condition we are thinking of;
the relation between the variables $u$ and $x$ is given by $u=F(x)$,
\begin{equation}
\label{eq:FFF}
F(x)=\frac{\sin\frac{\pi z_{41}}{x+z_{41}} \sin\frac{\pi z_{32}}{x+z_{41} }}{ \sin\frac{\pi z_{31}}{x+z_{41} }\sin\frac{\pi z_{42}}{x+z_{41} } }, 
\qquad
z_{ij}=z_i-z_j.
\end{equation}
In particular, if the boundary condition (b.c.) is the Neumann and $O_j(z)$ 
 is an on-shell tachyon $T_j(z)$ given by \eqref{eq:ttt},
the formula reduces to 
\begin{equation}
\begin{split}
&\int_a^b dx \langle B T_1(z_1) T_2(z_2)T_3(z_3) T_4(z_4)\rangle_{C_L} \\
=&\int_{F(a)}^{F(b)} du u^{-\alpha's-2}(1-u)^{-\alpha't-2}
\end{split}
\end{equation}
with $s$, $t$, $u$ the Mandelstam variables
\begin{equation}
s=-(k_1+k_2)^2,\quad t=-(k_1+k_3)^2,\quad u=-(k_1+k_4)^2,\quad 
\end{equation}
which, in the present case, satisfy
\begin{equation}
s+t+u=-\frac{4}{\alpha'}.
\end{equation}

\noindent
\subsection*{Derivation of the formula \eqref{eq:formula}}

\noindent
Consider the integrand of the  left hand side of \eqref{eq:formula}
\begin{equation}
\label{eq:aaaaaa}
\langle BO_1(z_1)  O_2(z_2) O_3(z_3) O_4(z_4) \rangle_{C_L}\equiv I(x).
\end{equation}
It is convenient to reduce this expression with \eqref{eq:4ptghostcrrtr}:  
\begin{equation}
\begin{split}
\left\langle V(z_1)V(z_2)V(z_3)V(z_4)\right\rangle^\text{matter}_{C_L}
 \frac{-{z_1}f_{234}+{z_2}f_{134}-{z_3}f_{124}+{z_4}f_{123}}{L}.
\end{split}
\end{equation}
Let us map this  from $C_L$ to UHP using  the conformal mapping
\begin{equation}
\xi=\tan \frac{\pi z}{x+z_4-z_1}
\end{equation}
to get the following expression\footnote{
It would be useful to notice that
\begin{equation}
\frac{d\xi}{dx}=-\left(\frac{d\xi}{dz}\right)\frac{z}{L}
\end{equation}
and
\begin{equation}
\frac{\partial u}{\partial \xi_i}=
\frac{(-1)^{i+1}}{(\xi_{31}\xi_{42})^2}
\prod_{k,l\ne i \text{ and }  k<l}
\xi_{lk}.
\end{equation}
} 
\begin{equation}
\label{eq:tiyuituit}
I(x)=
\langle  V_1(\xi_1)V_2(\xi_2)V_3(\xi_3)V_4(\xi_4) \rangle_{\text{UHP}}^\text{matter, b.c.}
\times 
\xi_{31}^2\xi_{42}^2\frac{du}{dx},
\end{equation}
where
\begin{equation}
u\equiv u(\xi_1,\xi_2,\xi_3,\xi_4)=\frac{\xi_{41}\xi_{32}}{\xi_{31}\xi_{42}}, \quad \xi_{ij}=\xi_i-\xi_j.
\end{equation}
We then exert the following SL$(2,\mathbb R)$ transformation on \eqref{eq:tiyuituit} 
\begin{equation}
\xi^\text{new}=\frac{(\xi-\xi_4)\xi_{23}}{(\xi-\xi_3)\xi_{24}}
\end{equation}
to obtain
\begin{equation}
I(x)=
\langle V_1(u) V_2(1)V_3(\infty) V_4(0) \rangle_{\text{UHP}}^\text{matter, b.c.}
\times 
\frac{du}{dx},
\end{equation}
where $ V_i(\infty)\equiv\lim_{\Lambda\to+\infty} \Lambda^2 V_i(\Lambda)$.
We therefore obtain that
\begin{equation}
\int_a^b dx I(x)=
\int_{F(a)}^{F(b)} du
\langle  V_1(u)  V_2(1) V_3(\infty) V_4(0) \rangle_{\text{UHP}}^\text{matter, b.c.} \quad (0\le a,\,b)
\end{equation}
with $F(x)$ given by \eqref{eq:FFF}.

\subsection{The main term}
Now, let us take the Erler-Maccaferri solution  
\begin{equation}
\label{eq:EMsolution}
\Psi=\Psi_T-{\Sigma} \Psi_T^\textrm{(b.c.)} \bar{\Sigma}, 
\end{equation}
where 
$\Psi_T^\textrm{(b.c.)}\in \mathcal H^{\textrm{(b.c.)}}$ is a tachyon vacuum solution of 
the open string field theory defined under the boundary condition (b.c.);
${\Sigma}$ and $\bar{\Sigma}$ are called \textit{regularized boundary condition changing operators}, characterized by the following relations:
\begin{equation}
\label{eq:algebra1}
\bar{\Sigma} {\Sigma}=1\in \mathcal H^\text{(b.c.)},
\end{equation}
\begin{equation}
Q_T{\Sigma}=Q_T \bar{\Sigma}=0,
\end{equation}
\begin{equation}
\label{eq:algebra3}
A_T{\Sigma}={\Sigma}A_T^\text{(b.c.)},\quad A_T^\text{(b.c.)}\bar{\Sigma}=\bar{\Sigma}A_T.
\end{equation}

\noindent
We here take $\Psi_T$ (and also $\Psi_T^\text{(b.c.)}$) to be the Schnabl solution \cite{Schnabl}  (see also \cite{Erler:2019nmz})
\begin{equation}
\label{eq:ssn}
\Psi_T=e^{\frac{K}{2}}c\frac{KB}{1-e^K}ce^{\frac{K}{2}}.
\end{equation}
The homotopy state
$A_T$ for the Schnabl solution 
is \cite{Ellwood+Schnabl}
\begin{equation}
 A_T=-\int_0^1 dx\, e^{xK}B.  
\end{equation}
From these materials, we can derive an simple expression for $W_\Psi$:
\begin{equation}
\label{eq:W-EM}
W_\Psi={\Sigma} e^{K} \bar{\Sigma}.
\end{equation}

\noindent
The external states $\{\mathcal O_j\}$, which satisfy 
\eqref{eq: psc1} and \eqref{eq: psc2}, are given as follows:
\begin{equation}
\label{eq:cohmology_general}
{\mathcal O}_j={\Sigma} O_j \bar{\Sigma}, 
\end{equation}
where $O_j $ is a $Q$-closed state given by \eqref{eq:ocv}.
In this subsection, we also assume an additional condition on $\{O_j\}$,
\begin{equation}
\label{eq:oooooosogooo}
O_iO_j=0, \quad i\ne j,
\end{equation}
which comes from the orthogonality condition \eqref{eq:orthogonality}.\\

\noindent
Now, we can express $H_{1234}$ in \eqref{eq:simplef2} as follows:
\begin{equation}
\label{eq:integ1}
\begin{split}
H_{1234}=&\int {\mathcal O}_1 W_\Psi{\mathcal O}_2 W_\Psi{\mathcal O}_3 W_\Psi{\mathcal O}_4 A_T\\
=&\int_0^1 dx \langle BO_1(-\tfrac{3}{2})O_2(-\tfrac{1}{2})O_3(\tfrac{1}{2}) O_4(\tfrac{3}{2})\rangle_{C_{3+x}} \\
=&\int_{0}^{\frac{1}{2}} du   
\langle  V_1(u)  V_2(1) V_3(\infty) V_4(0)\rangle_\text{UHP}^{\text{matter, (b.c.)}}  
\end{split}
\end{equation}
and
\begin{equation}
H_\Psi=\int_{-\infty}^{\infty} du   
\langle  V_1(u)  V_2(1) V_3(\infty) V_4(0)\rangle_\text{UHP}^{\text{matter, (b.c.)}}.
\end{equation}
Therefore, the main term of the gauge invariant quantity reproduces  
an expression for 
the four-point scattering amplitude of on-shell states
in the
first quantized theory. \\

\noindent
In particular, when (b.c.) is Neumann and $O_j(z)=T_j(z)$ with an additional condition $k_i \cdot k_j >-\frac{1}{2\alpha'}$,  
which comes from the orthogonality condition \eqref{eq:oooooosogooo}, $H_\Psi$ is given by
\begin{equation}
\label{eq:veneziano1}
H_\Psi=\int_{0}^{1} du u^{-\alpha's-2}(1-u)^{-\alpha't-2}+(s\leftrightarrow u)+(t\leftrightarrow u). 
\end{equation}
Of course, by analytic continuation with respect to $s$ and $t$, we can obtain the scattering amplitudes for general external momentum.
\\

\noindent
A couple of comments are in order here: 
\begin{itemize}
\item In the original construction \cite{ErlerMaccaferri},  ${\Sigma} \bar{\Sigma}$ is supposed to be a number which is the ratio of $g$-function $g_*/g_0$; 
but in the study of the theory around the classical solution \cite{Kishimoto:2014yea} or in the modified construction \cite{Erler_talk,EM2}, this product is a projector, 
\begin{equation}
({\Sigma} \bar{\Sigma})^2={\Sigma} \bar{\Sigma}. 
\end{equation}
Assuming this relation, we can prove that the elements of the cohomology of $Q_{\Psi}$ with the ghost number one can always be written in the form \eqref{eq:cohmology_general}. 

\item From this calculation, we can clearly see a meaning of the orthogonality condition \eqref{eq:orthogonality}; 
the region of the external momentum satisfying \eqref{eq:oooooosogooo} is given by
\begin{equation}
\label{eq:mitasu}
\{\alpha' s< -1\}\cap \{\alpha' t< -1\}\cap\{\alpha' u< -1\}.
\end{equation}
This is exactly the region where the integrals in \eqref{eq:veneziano1} converges.

\end{itemize}

\subsection{Contribution from the boundary term}
\noindent
This and the following subsection will focus on the case of  (b.c.)= Neumann and $O_j=T_j$  for simplicity, while 
the results can be generalized appropriately.\\

\noindent
As stated above, 
the expression \eqref{eq:veneziano1} diverges 
when the external momentums $\{k_j\}$ do not satisfy \eqref{eq:mitasu}. 
In such a case, the contribution from the boundary term $K_\Psi$ 
cancels this divergence precisely. 
To see how this works, let us write the contribution of $A_T-A_{\Psi}$ as:
\begin{equation}
\label{eq:formal}
\begin{split}
\bar{\Sigma}(A_T-A_\Psi){\Sigma}
&=-\int_0^1 dx e^{xK}B-\frac{B}{K}\\
&=-\lim_{\epsilon\to0}\left[\int_{\epsilon}^1 dx e^{xK}B+\frac{B}{K}e^{xK}|_{x=\epsilon}\right]\\
&\equiv \lim_{\epsilon\to0} \int_{\epsilon}^1 f_1(x)+f_0(x)|_{x=\epsilon}.
\end{split}
\end{equation}
It is important to notice that the integrand of the first term is given by 
acting $d$ on 
the integrand\footnote{Here we regard the second term as a zero-dimensional integral of $f_0(x)$. } of the second term:
\begin{equation}
d f_0(x) =f_1(x). 
\end{equation}
This relation should still hold after we put \eqref{eq:formal} inside the correlation function and make a change of integration variable from $x$ to $u$.  
In other words, \eqref{eq:integ1} receives the correction from the boundary term as follows:
\begin{equation}
\label{eq:tobusaki}
\lim_{\epsilon'\to 0}\int_{\epsilon'}^{\frac{1}{2}} du u^{-\alpha's-2}(1-u)^{-\alpha't-2}
 +g_0(u)|_{u=\epsilon'}
\end{equation}
with
\begin{equation}
\label{eq:bibun}
dg_0(u)=u^{-\alpha's-2}(1-u)^{-\alpha't-2} du.
\end{equation}
Since we will send $\epsilon'$ to zero at the end of this calculation, 
we can choose $g_0(u)$ as a function with finite terms
\begin{equation}
g_0(u)=\sum_{\alpha <-1}\frac{C_{\alpha}}{\alpha+1}u^{\alpha+1},
\end{equation}
where $C_\alpha$ are given by the Laurent coefficients of $dg_0(u)$ around $u=0$,\footnote{
We can drop the constant term in $g_0(u)$ for the following reason: 
after summing over all the terms, the zero-dimensional integral of $g_0(u)$ is over the boundary of 
union of intervals,  which is a closed zero-manifold. This zero-dimensional integral is then determined without ambiguity.
}
\begin{equation}
u^{-\alpha's-2}(1-u)^{-\alpha't-2}=\sum_{\alpha}C_{\alpha}u^\alpha. 
\end{equation}
This is nothing but the minimal subtraction of divergent terms of the main term and the result 
 \eqref{eq:veneziano1}
 is finally replaced by the well-known Veneziano amplitude with Euler's beta function. \\

\noindent
As mentioned earlier, this treatment of the boundary term is almost completely parallel to Sen's discussion \cite{Sen:2019jpm}. We therefore expect that several methods to evaluate the boundary term addressed in his paper will be useful 
 in practical calculations.

\subsection{Calculation for fat external states $\mathcal O_j^\text{fat}$ }

Up to this point, we have considered somewhat special external states $\mathcal{O}_j$,
 given by \eqref{eq:cohmology_general} with \eqref{eq:ocv}, 
in order to %
satisfy the orthogonality condition~\eqref{eq:orthogonality}.
In the following, we will demonstrate that the formula \eqref{eq:4pt} works well  
for a different type of external states which does not satisfy~\eqref{eq:orthogonality}.  \\

\noindent
As a most typical example, we consider the following \textit{fat} external states $\mathcal O^{\textrm{fat}}_j$: 
\begin{equation}
\label{eq:faaaaaat}
\mathcal O^{\textrm{fat}}_j={\Sigma}  O^{\textrm{fat}}_j \bar{\Sigma},
\end{equation}
with
\begin{equation}
\begin{split}
O^{\textrm{fat}}_j
&=e^{a_jK}O_j e^{a_jK}\\
&=e^{a_jK}e^{ik_j\cdot X}e^{a_jK},      \qquad a_j\ge 0 .  
\end{split}
\end{equation}
Again using the formula \eqref{eq:formula}, the main term can be reduced as follows:
first let us take a term $H_{1234}$ which can be expressed as 
\begin{equation}
\begin{split}
H_{1234}\equiv
&\int {\mathcal O^{\textrm{fat}}_1} W_\Psi{\mathcal O^{\textrm{fat}}_2} W_\Psi{\mathcal O^{\textrm{fat}}_3} W_\Psi{\mathcal O^{\textrm{fat}}_4} A_T\\
=&\int_{0+a_1+a_4}^{1+a_1+a_4} dx \langle B O_1(z_1) O_2(z_2) O_3(z_3) O_4(z_4)\rangle_{C_{L}} \\
=&\int_{A_{1234}}^{B_{1234}} du u^{-\alpha's-2}(1-u)^{-\alpha't-2},
\end{split}
\end{equation}
where
\begin{equation}
L=3+a_1+2(a_2+a_3)+a_4+x,
\end{equation}
\begin{equation}
\begin{split}
z_1=-a_2-a_3-a_4-\frac{3}{2}, &\quad
z_2=a_1-a_3-a_4-\frac{1}{2},\\
z_3=a_1+a_2-a_4+\frac{1}{2}, &\quad
z_4=a_1+a_2+a_3+\frac{3}{2},
\end{split}
\end{equation}
\begin{equation}
A_{1234}=F(a_1+a_4),\quad B_{1234}=F(1+a_1+a_4).
\end{equation}
Note the symmetric relation of $B_{ijkl}$, 
\begin{equation}
B_{4321}=B_{1234}, 
\end{equation}
and we can write
\begin{equation}
H_{1234}+H_{4321}
=
\int_{A_{1234}}^{1-A_{4321}} du u^{-\alpha's-2}(1-u)^{-\alpha't-2}. 
\end{equation}
Therefore, summing up over the permutation of indices, we obtain
\begin{equation}
H_\Psi=\frac{1}{4}\sum_{\textrm{C.P.}}\int_{A_{1234}}^{1-A_{4321}} du u^{-\alpha's-2}(1-u)^{-\alpha't-2}
+(s\leftrightarrow u)+(t\leftrightarrow u),
\end{equation}
where the summation on the right hand side is over the cyclic permutation (C.P.) of the indices. 
In this case, the domain of the integralation of $H_\Psi$ does not cover the entire moduli space. \\

\noindent
The missing region of the moduli space comes out from the boundary term~\eqref{eq:surface}. 
To see this, notice that with $g_0(u)$ defined in \eqref{eq:bibun} we can express the contribution from the boundary term as
\begin{equation}
\label{eq:H+K}
H_{1234}+K_{1234}=
\int_{A_{1234}}^{B_{1234}} du u^{-\alpha's-2}(1-u)^{-\alpha't-2}
 +g_0(u)\big|_{u=A_{1234}}. 
\end{equation}
The relation \eqref{eq:bibun} implies that 
we can replace $A_{1234}$ in \eqref{eq:H+K} with any positive number
$\epsilon'>0$ without changing the value of the right hand side:
\begin{equation}
H_{1234}+K_{1234}=
\int_{\epsilon'}^{B_{1234}} du u^{-\alpha's-2}(1-u)^{-\alpha't-2}
 +g_0(u)\big|_{u=\epsilon'}. 
\end{equation}
By sending $\epsilon'$ to zero, this equation reduces to \eqref{eq:tobusaki}. We will therefore be led to the same conclusion as before.

\subsection{On formal surface terms}
\label{sec:huhuihi}
Now, let us study a couple of delicate questions stated in Section~\ref{sec:manifestly} and Section~\ref{sec:formal}:  
each of them asks whether formal surface terms of the Witten integral vanish when their integrand has some factors of  
$A_\Psi$'s.  
Let us assume $\Psi=0$  in this subsection for simplicity.\footnote{
This subsection is essentially 
based on an old unpublished work of one of the authors with Yuji Okawa, which is 
partly presented in~\cite{Masuda:2012cj} or a talk at "String Field Theory 2011," Czech Academy of Sciences 
(\verb|https://sft11.fzu.cz/Talks/Masuda_sft2011.pdf|).
}  
\\

\noindent
On the one hand,  
a typical formal surface integral which appears in the proof in
 Section \ref{sec:manifestly} is
 \begin{equation}
\int Q((A_T-A_{0})\mathcal O_1 Y \mathcal O_2 W\mathcal O_3 W\mathcal O_4).
\end{equation}
This quantity is actually zero.  
Let us consider why it is the case in relation to the behavior of the following function as $x$ approaches the infinity,
\begin{equation}
\label{eq:integrated}
f(x)\equiv \int e^{xK}B\mathcal O_1 (QY) \mathcal O_2 W\mathcal O_3 W\mathcal O_4.
\end{equation}
If we sent $x$ to the infinity, $f(x)$ decreases as $O({x^{-1-\epsilon}})$ with some positive number $\epsilon>0$. This follows from the definition of $Y$,  
$Y=UA_TU^{-1}-A_T$. 
}\\

\noindent
Therefore, by integrating \eqref{eq:integrated}, we can write the following expression
\begin{equation}
\int \frac{B}{K}\mathcal O_1 (QY) \mathcal O_2 W\mathcal O_3 W\mathcal O_4
=\int_0^\infty dx f(x).
\end{equation}
In other words, we can express $A_\Psi$ as a superposition of the wedge states in this case. 
As a result, we can perform the partial integration as usual 
\begin{equation}
\begin{split}
&\int A\mathcal O_1 (QY)\mathcal O_2 W\mathcal O_3 W\mathcal O_4
=\int W\mathcal O_1 Y\mathcal O_2 W\mathcal O_3 W\mathcal O_4
\end{split}
\end{equation}
by discarding the surface integral.  
The same applies to  the other formal surface terms appeared in Section 4.2.\\

\noindent
An important corollary: the above discussion implies $I_{\Psi=0}^{(4)}$ is invariant under the replacement of $\Psi_T$ from Schnabl's solution to any other tachyon vacuum solution; in turn, since $I_\Psi^{(4)}$ is invariant under \eqref{eq:all_sim_transf} we can also conclude that \textit{$I_\Psi^{(4)}$ is invariant under the arbitrary gauge transformation around any classical solution $\Psi$.  }\\

\noindent
On the other hand, the formal expression \eqref{eq:bizarre} in this case is given by
\begin{equation}
\begin{split}
&\sum_{(i,j)} QR_{i,j}
=\int Q(A \mathcal O_1 A \mathcal O_2 W\mathcal O_3 W\mathcal O_4+...+
W \mathcal O_1 W \mathcal O_2 A\mathcal O_3 A\mathcal O_4).
\end{split}
\end{equation}
We have already seen that this quantity is not zero. 
This result can be understood by behavior of the following function:
\begin{equation}
f(x)\equiv \int e^{xK}B\mathcal O_1 W \mathcal O_2 W\mathcal O_3 W\mathcal O_4.
\end{equation}
This time, $xf(x)$ does not vanish as $x\to \infty$.
Therefore, $\frac{B}{K}$ cannot be expressed by a superposition of the wedge states, and the {formal} surface integral can be non-zero
without any contradiction:
\begin{equation}
\int (QA) \mathcal O_1 A \mathcal O_2 W\mathcal O_3 W\mathcal O_4 
\ne
-\int A \mathcal O_1 (QA) \mathcal O_2 W\mathcal O_3 W\mathcal O_4.
\end{equation}

\section{Concluding remarks}

So far, we have studied a series of new gauge invariant quantities in the open string field theory. 
Calculation with our new formula apparently does not require gauge fixing, which is necessitated to obtain a propagator in the Feynman rule.   
The conformal map used in our calculation can also be easier 
depending the choice of $\Psi_T$. 
\\

\noindent
We believe that our results are new and thought-provoking; yet, it would be more appropriate to say that we have excavated a lot of questions we have not understood yet, rather than we have 
clarified something. 
The open string field theory seems to have some simple and unexpected structure which we do not fully know yet.
\\

\noindent

\noindent
The results of this paper alone cannot be said to achieve the application purpose mentioned in the third paragraph of the introduction, and further research is required.
\noindent
In the following, we will make brief remarks on remaining questions, possible extensions and future works. 

\subsection{Basic questions on $I_\Psi^{(N)}$ left untouched}
\begin{enumerate}
\item To prove that the new formula $I_\Psi^{(N)}$ gives on-shell tree-level scattering amplitudes (hopefully, without assuming the Erler-Maccaferri solution).
\item To understand relation between $I_\Psi^{(N)}$ and the Feynman rule.
In particular, a calculation with $I_\Psi^{(N)}$ avoids the procedure of gauge fixing apparently, while our $A=A_T-A_\Psi$ (or $A_T$ alone) looks a variant of a propagator. 
\item To understand the redundancy of the definition of the formula. 
For example, the following formula also reproduces the on-shell four point amplitudes for the Erler-Maccaferri solution
\begin{equation}
\begin{split}
\sum_\sigma \int (1+W_\Psi)A\mathcal O_{\sigma(1)}(W_\Psi)^2\mathcal O_{\sigma(2)}(W_\Psi)^2\mathcal O_{\sigma(3)}(W_\Psi)^2\mathcal O_{\sigma(4)}.
\end{split}
\end{equation}
This formula produces a gauge invariant quantity. Such a redundancy seems to exist for general $N\ge 3$. 

\item To  study gauge invariant subsets of $I_\Psi^{(N)}$,  
 clarify its simplest unit and understand its meaning. 

\item To understand the nature of $W_\Psi$.  
The expression  \eqref {eq:W-EM} suggests that $W_\Psi$ is a fundamental quantity. 
It might be possible to use it to extract some basic property of the theory expanded around the classical solution, 
such as 
the boundary state or the spectrum of the theory (see also Ellwood~\cite{Ellwood:2009zf}).
\item To investigate the validity of $A_T$.  
The result of numerical calculation by~\cite{GI,I} (that is, the cohomology is not empty at non-standard ghost numbers) suggests that $ A_T $ may not be defined over the whole state space $\mathcal H$. 
This means 
 there is some room for doubt in \eqref{eq:htp}, which is one of the key formulas our discussion is based on. 

\item To fix the combinatorial constant $C_N$ in \eqref{eq:nten}. 
\end{enumerate}
\subsection{Possible extension to other  theories or settings}
\begin{enumerate}
\item To explore a similar formula in 
other string field theories 
including supersymmetric  or closed string field theories. 
Some of these extensions will follow if the relation between different string field theories are well understood \cite{Matsunaga:2019fnc, Erler-Matsunaga}. 
\item To write down and study a formula for the off-shell amplitudes.
\item To explore a similar formula for loop amplitudes.
\item To study a similar formula for on-shell disk amplitudes with closed string insertions. 
This may be related to a question how closed string physics can be described in open string field theory.

\end{enumerate}

\subsection{Problems related to numerical study}
\begin{enumerate}
\item We would like to evaluate $I_\Psi^{(N)}$ numerically, especially for numerical solutions to help to obtain their physical interpretation \cite{Kudrna:2018mxa,Kudrna:thesis, KSV}.

\item 
In order to perform the above calculation, it is necessary to solve 
\begin{equation}
Q_\Psi \mathcal O_j=0
\end{equation}
systematically to obtain the spectrum around the classical solution.
When a certain classical solution to \eqref{eq:eomoemoe} is given, it might be possible to obtain such information by sending  `the coupling constant' to small. 
\end{enumerate}

\subsection{Miscellaneous problems}
\begin{enumerate}
\item 
`What are the basic observables of the open string field theory' and
`what quantities can be read naturally from classical solutions  \cite {Ellwood, Kawano:2008ry, Kiermaier:2008qu,Kudrna:2012re, Kudrna:2014rya,thesisKB,BRR}' 
are important questions related to one another.  
In addition, there are physical quantities which are not defined yet even though they are expected to exist. 
To name a few, gauge invariant quantities obtained from the vacuum expectation value of the gauge field of the system \cite {Ishibashi:2016xak, Ishibashi:2018ynb} 
or what corresponds to the Wilson loop in local gauge field theory. 
\item It might be interesting to study  identity based solutions by using $I^{(N)}_\Psi$. 
There are many preceding studies to calculate physical quantities for the identity based solutions \cite{Inatomi:2012nv, Kishimoto:2013sra, Kishimoto:2014qza, Ishibashi:2014mua, Kishimoto:2014lua, Zeze:2010sr,Arroyo:2010sy}. 
As far as we choose the external states appropriately, 
$I_\Psi^{(N)}$ will be well-defined even if we take $\Psi$ or $\Psi_T$ from 
these identity based solutions. 
We would thus be able to calculate $I_\Psi^{(N)}$ without any special manipulation or regularization.

\item It may be suggestive that $ I_\Psi^{(N)}$ can be expressed as a formal surface integral \eqref{eq:bizarre}.
It reminds us the works of Ellwood~\cite {Ellwood:2009zf} or Erler-Maccaferri~\cite{Erler:2012qn}. The appearence of $ \frac {B}{K} $ reminds us an attempt of constructing lump solutions or multi-brane solutions \cite{Erler:2011tc, Murata:2011ex, Murata:2011ep, Hata:2011ke}. 
Furthermore, it also suggests that $I_\Psi$ is related to  an expression of an on-shell amplitude derived by Sen \cite{Sen:2019jpm}, which 
can be transformed into an integral of some function 
over a lower-dimensional subspace in the moduli space. 

\item[$3'$.] $I_\Psi^{(N)}$ might be related to the  winding number of  Hata and Kojita \cite{Hata:2011ke}. 
When the energy can be written as an integer multiple of something, the scattering amplitude is also expected to be so.

\item 

Although few people may be interested in \textit{fattening} the Ellwood invariant, the authors consider it an important issue that can have meaning in multiple contexts. 
This problem can be described as follows: 
\begin{enumerate}
\item Construct a non-trivial gauge invariant quantity from $\mathcal O^{(2)}\in \mathcal H$ which is an element of $Q_\Psi$-cohomology at ghost number two and a classical solution $\Psi$:
\begin{equation}
E_\text{fat}(\mathcal O^{(2)}, \Psi)
\end{equation}
which is invariant under 
\begin{equation}
\Psi                    \to U(\Psi+Q)U^{-1}
\end{equation}
and
\begin{equation}
\mathcal O^{(2)}\to  \mathcal O^{(2)}+Q_\Psi(...). 
\end{equation}
\item Prove that it is equivalent to the Ellwood invariant in some sense. 
\end{enumerate}
One obvious application is the energy calculation of the identity-based solutions.
In another context, the existence of these quantities would support the possibility of interpreting $\mathcal O^{(2)}$ as a closed string state. 

\item 
There is no doubt that Sen's work \cite{Sen:2019jpm} played an essential role in our calculations.
Meanwhile, whether the same results can be reproduced by using other prescriptions is also an important question. 
One possibility is to use something similar to the epsilon regularization, which has been studied to regularize  classical solutions \cite{Erler:2011tc, Murata:2011ex, Murata:2011ep, Hata:2011ke,Erler:2012qn,Erler:2012qr}. In fact,  perturbative calculations in \cite{Larocca:2017pbo} were performed in such a way.

\end{enumerate}


\noindent
\subsubsection*{Acknowledgement} 
\begin{itemize}
\item 
The authors would like to thank Ted Erler, Mat\v{e}j Kudrna and Yuji Okawa 
 for
valuable discussion or comments on their results. In particular, they would like to thank Carlo Maccaferri for 
useful discussion concerning the results of Section~5, 
for suggesting some topics treated in Section~6 
and for constructive comments on the draft. 
They would also like to thank Martin Schnabl for his suggestion to select  \eqref{eq:ssn} as a reference~$\Psi_T$ in Section 5 and for discussion (with M. Kudrna) concerning numerical studies.  
Special thanks also go to Albin James for providing the authors detailed and helpful comments on the draft and for finding several important typographical errors in the equations.
The authors would also like to thank  Renann Lipinski Jusinskas
for offering them concrete and constructive suggestions on the draft.
\item 
Some of these discussions were took place in the workshop ``String Theory from a worldsheet perspective,"  
and the final part of this work was done there. 
The authors would like to thank the organizers, the Galileo Galilei Institute for Theoretical Physics and INFN for hospitality and partial support during this workshop. 
TM would also like to thank 
the organizers of the conference ``String-Math 2019" for their hospitality. He also would like to thank participants of the conference 
who showed their interests on the present work. 
This experience motivated the authors to write a slightly long Section 2.

\item This work is supported by 
the Czech Science Foundation 
(GA\v{C}R)
 grant 17-22899S. 
%
The work of H.M. is also supported by Praemium Academiae and RVO: 67985840. 

\item The authors are deeply grateful to the wonderful musical environment of Prague. Such an environment has given them a mental state that is suitable for carrying out the present work. 
During this study,  two major music festivals \textit{the Rudolf 
Firku\v{s}n\'{y}
 piano festival 2018} and \textit{the Prague spring music festival 2019}  took place.  
They would like to thank the organizers and players involved, including Grigory Sokolov. 
TM would also like to thank: 
KBS Symphony Orchestra with Sunwoo Yekwon; 
Angela Hewitt; 
Making teams of 
\textit{Don Giovanni} and 
\textit{the Magic Flute} on the Estates Theatre; 
%
%
%
\v{S}imon Vose\v{c}ek, 
Prague Philharmonia and Ben Glassberg; 
 l'Orchestre National du Capitole de Toulouse 
 and Tugan Sokhiev;  
%
Sharon Kamm, Matan Porat, 
Josef \v{S}pa\v{c}ek 
and 
Petr Nouzovsk\'{y}; 
Members of the Philharminia Quartet. 

\end{itemize}


\noindent

\clearpage


\begin{thebibliography}{99}



  
  \bibitem{Witten} 
  E.~Witten,
  ``Noncommutative Geometry and String Field Theory,''
  Nucl.\ Phys.\ B {\bf 268} (1986) 253.
 
  \bibitem{Ellwood} 
  I.~Ellwood,
  ``The Closed string tadpole in open string field theory,''
  JHEP {\bf 0808} (2008) 063
  [arXiv:0804.1131 [hep-th]].

 
  \bibitem{Polchinski:1998rq}
  J.~Polchinski,
  ``String theory. Vol. 1: An introduction to the bosonic string,''
  Cambridge University Press, 1998.
  
  
\bibitem{GMW}
  S.~B.~Giddings, E.~J.~Martinec and E.~Witten,
  ``Modular Invariance in String Field Theory,''
  Phys.\ Lett.\ B {\bf 176} (1986) 362.

\bibitem{Z2}
  B.~Zwiebach,
  ``A Proof that Witten's open string theory gives a single cover of moduli space,''
  Commun.\ Math.\ Phys.\  {\bf 142} (1991) 193. 

\bibitem{Taylor:2003gn}
  W.~Taylor and B.~Zwiebach,
  ``D-branes, tachyons, and string field theory,''
  hep-th/0311017.

\bibitem{reviews_analytic_solution_0}
  E.~Fuchs and M.~Kroyter,
  ``Analytical Solutions of Open String Field Theory,''
  Phys.\ Rept.\  {\bf 502} (2011) 89
  [arXiv:0807.4722 [hep-th]].
  
  \bibitem{reviews_analytic_solution_1}
  Y.~Okawa,
  ``Analytic methods in open string field theory,''
  Prog.\ Theor.\ Phys.\  {\bf 128} (2012) 1001.

\bibitem{reviews_analytic_solution_2}
  M.~Schnabl,
  ``Algebraic solutions in Open String Field Theory - A Lightning Review,''
  Acta Polytechnica 50, no. 3 (2010) 102
  [arXiv:1004.4858 [hep-th]].

  \bibitem{Schnabl}  
  M.~Schnabl,
  ``Analytic solution for tachyon condensation in open string field theory,''
  Adv.\ Theor.\ Math.\ Phys.\  {\bf 10} (2006) no.4,  433
  [hep-th/0511286].

\bibitem{Okawa}
  Y.~Okawa,
  ``Comments on Schnabl's analytic solution for tachyon condensation in Witten's open string field theory,''
  JHEP {\bf 0604} (2006) 055
  [hep-th/0603159].

  
\bibitem{Schnabl:2002gg}
  M.~Schnabl,
  ``Wedge states in string field theory,''
  JHEP {\bf 0301} (2003) 004
  [hep-th/0201095].

\bibitem{Rastelli:2000iu}
  L.~Rastelli and B.~Zwiebach,
  ``Tachyon potentials, star products and universality,''
  JHEP {\bf 0109} (2001) 038
  [hep-th/0006240].


\bibitem{LPP1}
  A.~LeClair, M.~E.~Peskin and C.~R.~Preitschopf,
  ``String Field Theory on the Conformal Plane. 1. Kinematical Principles,''
  Nucl.\ Phys.\ B {\bf 317} (1989) 411.

\bibitem{LPP2}
  A.~LeClair, M.~E.~Peskin and C.~R.~Preitschopf,
  ``String Field Theory on the Conformal Plane. 2. Generalized Gluing,''
  Nucl.\ Phys.\ B {\bf 317} (1989) 464.


 
  \bibitem{ErlerMaccaferri}
  T.~Erler and C.~Maccaferri,
  ``String Field Theory Solution for Any Open String Background,''
  JHEP {\bf 1410} (2014) 029
  [arXiv:1406.3021 [hep-th]].


\bibitem{Kato:1982im}
  M.~Kato and K.~Ogawa,
  ``Covariant Quantization of String Based on BRS Invariance,''
  Nucl.\ Phys.\ B {\bf 212} (1983) 443.

\bibitem{Sen:1999xm}
  A.~Sen,
  ``Universality of the tachyon potential,''
  JHEP {\bf 9912} (1999) 027
  [hep-th/9911116].



\bibitem{Erler:2009uj}
  T.~Erler and M.~Schnabl,
  ``A Simple Analytic Solution for Tachyon Condensation,''
  JHEP {\bf 0910} (2009) 066
  [arXiv:0906.0979 [hep-th]].
 





\bibitem{HI1} 
  A.~Hashimoto and N.~Itzhaki,
  ``Observables of string field theory,''
  JHEP {\bf 0201} (2002) 028
  [hep-th/0111092].

\bibitem{GRSZ1}
  D.~Gaiotto, L.~Rastelli, A.~Sen and B.~Zwiebach,
  ``Ghost structure and closed strings in vacuum string field theory,''
  Adv.\ Theor.\ Math.\ Phys.\  {\bf 6} (2003) 403
  [hep-th/0111129].

\bibitem{Z1}
  B.~Zwiebach,
  ``Interpolating string field theories,''
  Mod.\ Phys.\ Lett.\ A {\bf 7} (1992) 1079
  [hep-th/9202015].

 \bibitem{Kudrna:2012re}
  M.~Kudrna, C.~Maccaferri and M.~Schnabl,
  ``Boundary State from Ellwood Invariants,''
  JHEP {\bf 1307} (2013) 033
  [arXiv:1207.4785 [hep-th]].




\bibitem{Baba:2012cs}
  T.~Baba and N.~Ishibashi,
  ``Energy from the gauge invariant observables,''
  JHEP {\bf 1304} (2013) 050
  [arXiv:1208.6206 [hep-th]].










 
\bibitem{Sen-Zwiebach}
  A.~Sen and B.~Zwiebach,
  ``Tachyon condensation in string field theory,''
  JHEP {\bf 0003} (2000) 002
  [hep-th/9912249].


\bibitem{Takahashi:2002ez}
  T.~Takahashi and S.~Tanimoto,
  ``Marginal and scalar solutions in cubic open string field theory,''
  JHEP {\bf 0203} (2002) 033
  [hep-th/0202133].
 
  
\bibitem{Ellwood+Schnabl}
  I.~Ellwood and M.~Schnabl,
  ``Proof of vanishing cohomology at the tachyon vacuum,''
  JHEP {\bf 0702} (2007) 096
  [hep-th/0606142].


\bibitem{EFHM}
  I.~Ellwood, B.~Feng, Y.~H.~He and N.~Moeller,
  ``The Identity string field and the tachyon vacuum,''
  JHEP {\bf 0107} (2001) 016
  [hep-th/0105024].
 
\bibitem{Inatomi:2011xr}
  S.~Inatomi, I.~Kishimoto and T.~Takahashi,
  ``Homotopy Operators and One-Loop Vacuum Energy at the Tachyon Vacuum,''
  Prog.\ Theor.\ Phys.\  {\bf 126} (2011) 1077
  [arXiv:1106.5314 [hep-th]].

\bibitem{Inatomi:2011an}
  S.~Inatomi, I.~Kishimoto and T.~Takahashi,
  ``Homotopy Operators and Identity-Based Solutions in Cubic Superstring Field Theory,''
  JHEP {\bf 1110} (2011) 114
  [arXiv:1109.2406 [hep-th]].
 
 
 
 
 
\bibitem{GI}
  S.~Giusto and C.~Imbimbo,
  ``Physical states at the tachyonic vacuum of open string field theory,''
  Nucl.\ Phys.\ B {\bf 677} (2004) 52
  [hep-th/0309164].
 
 
\bibitem{I}
  C.~Imbimbo,
  ``The Spectrum of open string field theory at the stable tachyonic vacuum,''
  Nucl.\ Phys.\ B {\bf 770} (2007) 155
  [hep-th/0611343].

 
 
\bibitem{Ellwood:2009zf}
  I.~Ellwood,
  ``Singular gauge transformations in string field theory,''
  JHEP {\bf 0905} (2009) 037
  [arXiv:0903.0390 [hep-th]].
   
\bibitem{Hene?} 
M. Henneaux, “Quantization of gauge systems,”  Princeton University Press (1992).
  
 
 
\bibitem{KO1}
T. Kugo and I. Ojima, “Manifestly Covariant Canonical Formulation of Yang-Mills Field Theories: Physical State Subsidiary Conditions and Physical S Matrix Unitarity,” Phys. Lett. B \textbf{73} (1978) 459.

\bibitem{KO2}
T. Kugo and I. Ojima, “Local Covariant Operator Formalism of Nonabelian Gauge Theories and Quark Confinement Problem,” Prog. Theor. Phys. Suppl. \textbf{66} (1979) 1.


\bibitem{Sen:2019jpm}
  A.~Sen,
  ``String Field Theory as World-sheet UV Regulator,''
  arXiv:1902.00263 [hep-th].

\bibitem{Erler:2012qn}
  T.~Erler and C.~Maccaferri,
  ``Connecting Solutions in Open String Field Theory with Singular Gauge Transformations,''
  JHEP {\bf 1204} (2012) 107
  [arXiv:1201.5119 [hep-th]].

\bibitem{Erler:2019nmz}
  T.~Erler, C.~Maccaferri and R.~Noris,
  ``Taming boundary condition changing operator anomalies with the tachyon vacuum,''
  JHEP {\bf 1906} (2019) 027
  [arXiv:1901.08038 [hep-th]].


\bibitem{Kishimoto:2014yea}
  I.~Kishimoto, T.~Masuda, T.~Takahashi and S.~Takemoto,
  ``Open String Fields as Matrices,''
  PTEP {\bf 2015} (2015) no.3,  033B05
  [arXiv:1412.4855 [hep-th]].


\bibitem{Erler_talk}
  T.~Erler,
 Talk at 
``String Field Theory and String Phenomenology," Harish-Chandra Research Institute, 14 Feb. 2018, \\
 \verb|http://www.hri.res.in/~strings/erler.pdf|\\
C. Maccaferri, 
Talk at 
  ``New Frontiers in String Theory," Yukawa Institute of Theoretical Physics, 19 July 2018.\\
  \verb|http://www2.yukawa.kyoto-u.ac.jp/~nfst2018/Slide/Maccaferri.pdf|  


\bibitem{EM2}
  T.~Erler, C.~Maccaferri,
  \textit{to appear.} 




\bibitem{Masuda:2012cj}
  T.~Masuda,
  ``Comments on new multiple-brane solutions based on Hata-Kojita duality in open string field theory,''
  JHEP {\bf 1405} (2014) 021
  [arXiv:1211.2649 [hep-th]].










\bibitem{Matsunaga:2019fnc}
  H.~Matsunaga,
  ``Light-cone reduction of Witten's open string field theory,''
  JHEP {\bf 1904} (2019) 143
  [arXiv:1901.08555 [hep-th]].

\bibitem{Erler-Matsunaga}
  T. Erler and H.~Matsunaga,
 \textit{to appear}. 

\bibitem{Kudrna:2018mxa}
  M.~Kudrna and M.~Schnabl,
  ``Universal Solutions in Open String Field Theory,''
  arXiv:1812.03221 [hep-th].

\bibitem{Kudrna:thesis}
  M.~Kudrna,
 Ph.D. thesis, Charles University in Prague, 2019.

\bibitem{KSV}
  M.~Kudrna, M.~Schnabl and J.~Vošmera,
  \textit{to appear}.


 \bibitem{Kawano:2008ry}
  T.~Kawano, I.~Kishimoto and T.~Takahashi,
  ``Gauge Invariant Overlaps for Classical Solutions in Open String Field Theory,''
  Nucl.\ Phys.\ B {\bf 803} (2008) 135
  [arXiv:0804.1541 [hep-th]].


\bibitem{Kiermaier:2008qu}
  M.~Kiermaier, Y.~Okawa and B.~Zwiebach,
  ``The boundary state from open string fields,''
  arXiv:0810.1737 [hep-th].


\bibitem{Kudrna:2014rya}
  M.~Kudrna, M.~Rap\v{c}\'{a}k and M.~Schnabl,
  ``Ising model conformal boundary conditions from open string field theory,''
  arXiv:1401.7980 [hep-th].


\bibitem{thesisKB}
  K. Budzik,
 Master's thesis, 
 Perimeter Institute for Theoretical Physics,
 2019.

\bibitem{BRR}
 K. Budzik, M. Rap\v{c}\'{a}k and J. M. Rojas, 
   \textit{to appear}.




\bibitem{Ishibashi:2016xak}
  N.~Ishibashi, I.~Kishimoto and T.~Takahashi,
  ``String field theory solution corresponding to constant background magnetic field,''
  PTEP {\bf 2017} (2017) no.1,  013B06
  [arXiv:1610.05911 [hep-th]].
 
 \bibitem{Ishibashi:2018ynb}
  N.~Ishibashi, I.~Kishimoto, T.~Masuda and T.~Takahashi,
  ``Vector profile and gauge invariant observables of string field theory solutions for constant magnetic field background,''
  JHEP {\bf 1805} (2018) 144
  [arXiv:1804.01284 [hep-th]].





\bibitem{Inatomi:2012nv}
  S.~Inatomi, I.~Kishimoto and T.~Takahashi,
  ``Tachyon Vacuum of Bosonic Open String Field Theory in Marginally Deformed Backgrounds,''
  PTEP {\bf 2013} (2013) 023B02
  [arXiv:1209.4712 [hep-th]].

\bibitem{Kishimoto:2013sra}
  I.~Kishimoto and T.~Takahashi,
  ``Gauge Invariant Overlaps for Identity-Based Marginal Solutions,''
  PTEP {\bf 2013} (2013) 093B07
  [arXiv:1307.1203 [hep-th]].

\bibitem{Kishimoto:2014qza}
  I.~Kishimoto and T.~Takahashi,
  ``Comments on observables for identity-based marginal solutions in Berkovits' superstring field theory,''
  JHEP {\bf 1407} (2014) 031
  [arXiv:1404.4427 [hep-th]].

\bibitem{Ishibashi:2014mua}
  N.~Ishibashi,
  ``Comments on Takahashi-Tanimoto's scalar solution,''
  JHEP {\bf 1502} (2015) 168
  [arXiv:1408.6319 [hep-th]].
 
 \bibitem{Kishimoto:2014lua}
  I.~Kishimoto, T.~Masuda and T.~Takahashi,
  ``Observables for identity-based tachyon vacuum solutions,''
  PTEP {\bf 2014} (2014) no.10,  103B02
  [arXiv:1408.6318 [hep-th]].


\bibitem{Zeze:2010sr}
  S.~Zeze,
  ``Regularization of identity based solution in string field theory,''
  JHEP {\bf 1010} (2010) 070
  [arXiv:1008.1104 [hep-th]].

\bibitem{Arroyo:2010sy}
  E.~A.~Arroyo,
  ``Comments on regularization of identity based solutions in string field theory,''
  JHEP {\bf 1011} (2010) 135
  [arXiv:1009.0198 [hep-th]].
  


\bibitem{Erler:2011tc}
  T.~Erler and C.~Maccaferri,
  ``Comments on Lumps from RG flows,''
  JHEP {\bf 1111} (2011) 092
  [arXiv:1105.6057 [hep-th]].
  
  

\bibitem{Murata:2011ex}
  M.~Murata and M.~Schnabl,
  ``On Multibrane Solutions in Open String Field Theory,''
  Prog.\ Theor.\ Phys.\ Suppl.\  {\bf 188} (2011) 50
  [arXiv:1103.1382 [hep-th]].

\bibitem{Murata:2011ep}
  M.~Murata and M.~Schnabl,
  ``Multibrane Solutions in Open String Field Theory,''
  JHEP {\bf 1207} (2012) 063
  [arXiv:1112.0591 [hep-th]].

  
  
  \bibitem{Hata:2011ke}
  H.~Hata and T.~Kojita,
  ``Winding Number in String Field Theory,''
  JHEP {\bf 1201} (2012) 088
  [arXiv:1111.2389 [hep-th]].

\bibitem{Erler:2012qr}
  T.~Erler and C.~Maccaferri,
  ``The Phantom Term in Open String Field Theory,''
  JHEP {\bf 1206} (2012) 084
  [arXiv:1201.5122 [hep-th]].




\bibitem{Larocca:2017pbo}
  P.~V.~Larocca and C.~Maccaferri,
  ``BCFT and OSFT moduli: an exact perturbative comparison,''
  Eur.\ Phys.\ J.\ C {\bf 77} (2017) no.11,  806
  [arXiv:1702.06489 [hep-th]].

 
\end{thebibliography}
\end{document}